\documentclass[twocolumn,times]{aastex631}
\usepackage{amssymb}
\received{\today}
\revised{}
\accepted{}
\submitjournal{ApJ}

\usepackage{color,comment}
\usepackage[normalem]{ulem}
\usepackage{newtxtext,newtxmath}
\usepackage{amsmath} 
\usepackage{ctable}

\newcommand{\angstrom}{\textup{\AA}}

\newcommand{\qpah}[1]{$q_{\rm PAH}$ }
\newcommand{\lpah}[1]{$L_{\rm PAH}$ }
\newcommand{\lfir}[1]{$L_{\rm FIR}$ }

\shorttitle{Modeling PAHs in Galaxy Simulations}
\shortauthors{Desika and the PAH Patrol}

\begin{document}

\title[]{A Framework for Modeling Polycyclic Aromatic Hydrocarbon Emission in Galaxy Evolution Simulations}

\correspondingauthor{Desika Narayanan}
\email{desika.narayanan@ufl.edu}

\author[0000-0002-7064-4309]{Desika Narayanan}
\affil{Department of Astronomy, University of Florida, 211 Bryant Space Sciences Center, Gainesville, FL 32611 USA}
\affil{University of Florida Informatics Institute, 432 Newell Drive, CISE Bldg E251, Gainesville, FL 32611}
\affil{Cosmic Dawn Center at the Niels Bohr Institute, University of Copenhagen and DTU-Space, Technical University of Denmark}

\author[0000-0003-1545-5078]{J.-D. T. Smith}
\affil{Ritter Astrophysical Research Center, Department of Physics and Astronomy,\\ University of Toledo, Toledo, OH 43606, USA}

\author[0000-0001-7449-4638]{Brandon S. Hensley}
\affiliation{Department of Astrophysical Sciences,  Princeton University, Princeton, NJ 08544, USA}

\author{Qi Li}
\affil{Max Planck Institute for Astrophysics, Garching bei Munchen, Germany}

\author[0000-0002-9235-3529]{Chia-Yu Hu}
\affil{Department of Astronomy, University of Florida, 211 Bryant Space Sciences Center, Gainesville, FL 32611 USA}

\author[0000-0002-4378-8534]{Karin Sandstrom}
\affil{Center for Astrophysics and Space Sciences, Department of Physics, University of California, San Diego,\\ 9500 Gilman Drive, La Jolla, CA 92093, USA}

\author[0000-0002-5653-0786]{Paul Torrey}
\affil{Department of Astronomy, University of Florida, 211 Bryant Space Sciences Center, Gainesville, FL 32611 USA}

\author[0000-0001-8593-7692]{Mark Vogelsberger}
\affil{Department of Physics, Kavli Institute for Astrophysics and Space Research, Massachusetts Institute of Technology, Cambridge, MA 021\
39, USA}

\author[0000-0003-3816-7028]{Federico Marinacci}
\affil{Department of Physics and Astronomy "Augusto Righi", University of Bologna, via Gobetti 93/2, 40129, Bologna, Italy}

\author[0000-0002-3790-720X]{Laura V. Sales}
\affiliation{Department of Physics and Astronomy, University of California, Riverside, CA, 92521, USA}




\begin{abstract}
We present a new methodology for simulating mid-infrared emission from
polycyclic aromatic hydrocarbons in galaxy evolution simulations.  To
do this, we combine theoretical models of PAH emission features as
they respond to varying interstellar radiation fields, grain size
distributions, and ionization states with a new model for dust
evolution in galaxy simulations.  We apply these models to $3$
idealized {\sc arepo} galaxy evolution simulations within the {\sc
  smuggle} physics framework.  We use these simulations to develop
numerical experiments investigating the buildup of PAH masses and
luminosities in galaxies in idealized analogs of the Milky Way, a
dwarf galaxy, and starburst disk.  Our main results follow.  Galaxies
with high specific star formation rates have increased feedback energy
per unit mass, and are able to efficiently shatter grains, driving up
the fraction of ultra small grains.  At the same time, in our model
large radiation fields per unit gas density convert aliphatic grains
into aromatics.  The fraction of dust grains in the form of PAHs
($q_{\rm PAH}$) can be understood as a consequence of these processes,
and in our model PAHs form primarily from interstellar processing
(shattering) of larger grains rather than from the growth of smaller
grains.  We find that the hardness of the radiation field plays a
larger role than variations in the grain size distribution in setting
the total integrated PAH luminosities, though cosmological simulations
are necessary to fully investigate the complex interplay of processes
that drive PAH band luminosities in galaxies.  Finally, we highlight
feature PAH strength variations, cautioning against the usage of
emission templates with constant feature strength ratios.

\end{abstract}




\section{Introduction}
\label{section:introduction}
The mid-infrared wavelength regime of galaxy spectral energy
distributions (SEDs) is dominated by a series of strong emission
features at wavelengths $\lambda=3.3-17 \micron$.  These features,
first observed by \citet{gillett73a} and \citet{merrill75a}, were
attributed to vibrational modes of polycyclic aromatic hydrocarbons (PAHs) in
the interstellar medium (ISM) of galaxies by \citet{leger84a} and \citet{allamandola85a}.
In this picture, the emission features originate in
ultrasmall dust grains (typically $<1000$ carbon atoms) that are
arranged chemically in aromatic rings \citep{draine01a}. These PAH
molecules stochastically absorb ultraviolet photons in single-photon
heating events, become highly excited, and then cool by emitting a
series of infrared (IR) photons via vibrational transitions.  Here, different types of bending
modes in the C-C and C-H molecular skeletons drive the individual
emission features between $3-17 \micron$ (see \citealt{tielens08a}, \citealt{armus20a}, and
\citealt{li20b} for reviews).

PAH emission is nearly ubiquitous in the spectra of star-forming galaxies.
\citet{helou00a} observed bright PAH emission in a sample of $7$
nearby galaxies with the Infrared Space Observatory
\citep[ISO;][]{kessler96a}, while \citet{smith07a} expanded on this
significantly as a part of the Spitzer Infrared Nearby Galaxies Survey
\citep{kennicutt03b}, and investigated PAH emission from $59$ nearby
galaxies.  A key finding from these studies is that PAH emission can
constitute as much as $10-20\%$ of the total infrared luminosity in a
galaxy \citep{smith07a,dale09a,lai20a}, though the relative feature
strengths can vary substantially both within and between sources
\citep{peeters04b}.  Of the individual PAH features, emission at $7.7
\micron$ dominates, contributing up to $\sim 40\%$ of the total PAH
luminosity \citep{hunt10a,wu10b,shipley13a}.

The prevalence of PAH emission features in the mid-IR regime of galaxy
SEDs coupled with their UV-based heating mechanisms have prompted a
number of authors to investigate the utility of PAHs as a tracer of
galaxy star formation rates
\citep{peeters04a,bendo08a,shipley16a,maragkoudakis18a,lai20a,whitcomb20a,evans22a}.
\citet{shipley16a} compiled $>100$ Spitzer Infrared Spectrograph (IRS)
detections of galaxies at $z<0.4$ and derived a linear relationship
between the H$\alpha$-measured star formation rate (SFR) of galaxies and their PAH
luminosities.  Similarly, the $7.7 \micron$ feature, which redshifts
into the $24 \micron$ MIPS filter on Spitzer, has been used in
numerous studies to determine the SFR of $z\sim 2$ galaxies
\citep{reddy06a,yan07a,siana09a,wuyts11b,rujopakarn13a}.  More recent
analysis of Spitzer IRS observations by \citet{whitcomb22a} and \citet{zhang22a}, as well as WISE photometry by \citet{chown21a} found that mid-IR PAH
features may trace galaxy molecular gas as well.

At the same time, some physical conditions may suppress the
PAH luminosities from galaxies ($L_{\rm PAH}$), causing deviations from this linear increase
of \lpah \ with galaxy SFR.  For example, there is
significant observational evidence that the PAH luminosity from galaxies is
suppressed in low-metallicity environments
\citep{wu06a,smith07a,engelbracht08a,hao09a,hunt10a,sandstrom10a,shivaei17a,aniano20a,shivaei22a}.
\citet{sandstrom10a} and \citet{chastenet19a} presented maps of the
Small Magellanic Cloud (SMC) and Large Magellanic Cloud (LMC),
demonstrating that PAH mass fractions increase with increasing galaxy metallicity.
\citet{aniano20a} compiled data from $53$ nearby galaxies and
derived fitting relationships between the PAH mass fraction,
and O/H abundances in the ISM.  While the origin of such a relationship is
unclear, it may result from photodestruction of PAHs in unshielded
environments \citep{voit92a,hunt10a,madden06a}, increased erosion via
thermal sputtering in a hotter ISM \citep{hunt11a}, a lack of seed
metals to grow into small grains, and/or a lack of seed dust grains that
shatter into smaller particles \citep{seok14a}.  Some authors have
further argued that the primary physical correlation may be with the
radiation field hardness and not metallicity
\citep[e.g.][]{gordon08a}, though it is possible that the smallest
grains are able to survive even in harsh radiation environments
\citep{lai20a}.

Similarly, PAH emission is observed to be weaker than otherwise
expected (given galaxy SFRs) in the vicinity of AGN \citep[see the
  recent review by][]{sajina22a}.  This has been observed both in
resolved imaging of nearby Seyfert nuclei or low-ionization nuclear
emission line regions (LINERs),
\citep[e.g.][]{smith07a,diamondstanic10a,sales10a}, as well as in unresolved
observations of galaxies at both low and high-redshift
\citep{rigopoulou99a,desai07a,farrah07a,pope08a,odowd09a}, though at
least for some AGN-hosting galaxies, this trend is unclear \citep{lai22a}.  The putative
physical mechanism behind the lack of PAH emission near AGN is
radiative destruction of ultra small grains \citep{voit92a,genzel98a}. Recent JWST observations of HII regions in nearby galaxies have demonstrated an anti-correlation between PAH fraction and ionization parameter, which may be in support of this scenario \citep{egorov22a}. 
 Other
works have suggested shocks as the main destruction mechanism for PAHs
near AGN \citep{zhang22a}.
At the same time, some studies have suggested that ultraviolet radiation
from AGN can actually excite PAH molecules \citep{howell07a,jensen17a}. 

 From a theoretical standpoint PAHs not only trace the physical
properties of galaxies \citep[e.g.][]{maragkoudakis22a}, but also
drive the evolution of the ISM.  For example, \citet{bakes94a}
determined that these ultra small grains can dominate the
photoelectric component to ISM heating in neutral gas, while at the same time PAHs
may be an important constituent of interstellar chemical reactions
\citep{lepp88a,bakes98a,weingartner01a}.  PAHs may be an important
catalyst for molecular hydrogen formation as well
\citep{thrower12a,foley18a}, and indeed have been detected even in CO dark molecular gas \citep{mcguire21a}.

Since their discovery, two broad approaches have emerged in modeling
the emission features of astrophysical PAHs: empirical models, and density functional theory calculations (that are sometimes combined with laboratory measurements).  \citet{draine01a} and
\citet{draine07a} pioneered the development of empirical models for
heating of ultrasmall aromatic carbonaceous grains -- and their emergent
emission features -- assuming a \citet{mathis83a}-like interstellar
radiation field.
\citet{draine21a} significantly broadened this model by considering
both a diverse range of incident radiation field spectral shapes, as well as bulk variations in the distributions of grain sizes and the ionization state.  These models are agnostic about the composition of PAHs themselves, but are designed in the context of reproducing a broader background of
observations, including infrared SEDs from galaxies and extinction
properties \citep[e.g.,][]{hensley20}.   Specifically, the cross sections, width and locations of the PAH bands that are modeled by \citet{draine21a} are tuned to match astrophysical spectra of galaxies, including the SINGS sources from \citet{smith07a}. In contrast, density functional theory modeling  involves computing the theoretical emission spectra for
grains of a diverse range of chemical compositions
\citep{bauschlicher10a,boersma14a,bauschlicher18a, mattioda20a,kerkeni22a,rigopoulou21a,vats22a}.   These ``database" approaches employ fitting techniques in order to deduce the PAH size distributions and ionization fractions from observations \citep[e.g.][]{maragkoudakis22a}.

With the launch of the JWST in 2021, observational studies of PAHs in
galaxies are poised to enter their renaissance -- already JWST is revealing the impact of galactic environment on PAH emission in galaxies near and far at unprecedented sensitivity and spatial resolution \citep[e.g.][]{armus22a,chastenet22a,dale22a,egorov22a,evans22a,langeroodi22a,lai22a,u22a,chastenet23a,sandstrom23a}.
What is missing thus far is quantitative methodology for modeling the physical processes that drive the evolution of dust in the interstellar medium in galaxies that span a wide range of physical conditions, and connecting these processes to the emergent PAH emission. 

The purpose of this paper is
to do just that.  In what follows, we develop a model for simulating
the mid-infrared emission from PAHs in hydrodynamic galaxy evolution
simulations.  Our paper is organized as follows: in
\S~\ref{section:methods}, we describe our methodology of simulating
the evolution of dust on the fly in galaxy evolution simulations, and
the coupling of this model with the \citet{draine21a} theoretical
model for the strength of PAH emission from grains in varying ISM
physical conditions.  In \S~\ref{section:galaxy_evolution} we
introduce a sample of $3$ galaxy evolution models that we will use for
the purposes of numerical experiments, and describe their physical
properties.  In \S~\ref{section:demonstration_of_methods}, we
demonstrate the results of our methodology by simulating the mid-IR SEDs and
images of our model galaxies.  In \S~\ref{section:masses_and_luminosities}, we explore the
origin of PAH masses and luminosities in these idealized galaxy
evolution simulations.  In \S~\ref{section:discussion}, we provide
discussion, and in \S~\ref{section:conclusions} we summarize.

\section{Model Implementation}

\label{section:methods}

\subsection{High-Level Overview and Modeling Philosophy}
Our main goal in this paper is to develop a modular framework for simulating PAH emission in galaxy evolution simulations.  To do so requires a multi-scale methodology with a large dynamic range of physical processes modeled.   While many of the methods that we include as subresolution processes are uncertain in the literature, our aim is to develop this model in a sufficiently parameterized way that new physics can be implemented or updated as the field evolves rapidly during the era of JWST.

The dominant drivers of the PAH spectrum in our model are: (i) the
dust grain size distribution; (ii) the intensity and spectrum of the interstellar
radiation field (ISRF); and (iii) the ionization fraction of the
PAHs.  In order to simulate the PAH emission spectrum from galaxies,
then, we must model each of these physical processes, and tie them
together.  To do this, we couple the \citet{draine21a} model -- that
describes the emitted PAH spectrum as a function of the grain size
distribution, ISRF, and ionization state -- with a new generation of
hydrodynamic galaxy evolution simulations that explicitly model the
formation, growth, and destruction processes of dust with a carefully tracked
distribution of sizes. In the following sub-sections, we describe the
details of each of these elements in turn.

\subsection{\texorpdfstring{Computing the PAH Emission Spectrum with Varying Physical Conditions: Summary of the \citet{draine21a} Model}{Computing the PAH Emission Spectrum with Varying Physical Conditions: Summary of the Draine et al. 2021 Model}}
We first begin with a summary of the \citet{draine21a} model for PAH
emission in a range of environments: this will set the stage for the
subsequent elements of our model.  In short, the \citet{draine21a}
model updates that of \citet{draine07a} in computing the sensitivity
of the PAH emission spectrum to the three major physical inputs: the
spectrum of the illuminating radiation field, the dust grain size
distribution and the PAH ionization state.

The \citeauthor{draine21a} model considers the starlight spectrum from
$14$ different radiation fields\footnote{Formally, these were actually computed twice: once in an unreddened mode, and a second time through a slab of dust with $A_{\rm V}=2$ to simulate progressive reddening in dusty clouds.}, all treated as single-age stellar
populations (SSPs): these radiation fields span a diverse range of
spectral shapes.  In detail, the SSP models are comprised of:
\citet{bruzual03a} starburst models ranging in range from $0.03-1$
Gyr; {\sc bpass} binary star SSP models \citep{eldridge17a,stanway18a}
models over the same age ranges, as well as a low metallicity model; and finally
an older stellar population akin to the bulge of M31
\citep{groves12a}.   Additionally a modified \citet{mathis83a} solar neighborhood-like model is included. 
 Because our models explicitly compute the impact of interstellar
reddening via a combination of on-the-fly dust evolution models
(\ref{section:dust_methods}) as well as full 3D Monte Carlo radiative transfer
(\ref{section:isrf}), we employ only the unreddened versions of these
stellar radiation fields from \citet{draine21a}.

This starlight intensity is parameterized by the rate of energy absorption onto grains given by the dimensionless intensity parameter:
\begin{equation}
  \label{equation:intensity_parameter}
U \equiv \frac{\int d\nu \ u_{*,\nu} c C_{\rm abs}\left(\nu\right)}{h_{\rm ref}} = \gamma_* \frac{u_*}{u_{\rm mMMP}}
\end{equation}
Here, $u_*$ is the energy density of the radiation field, $C_{\rm
  abs}$ is the orientation-averaged absorption cross section for a
standard dust grain \citep[here, the ``astrodust'' grain with porosity
  $P=0.2$ and radius $a_{\rm eff}=0.1 \mu$m;][]{hensley22a}, and
$h_{\rm ref}$ is the heating rate for the modified \citet{mathis83a}
spectrum (hereafter, the mMMP spectrum) where $h_{\rm ref} = 1.958
\times 10^{-12}$ erg/s.  $\gamma_*$ is a dust absorption weighted radiation field energy density, which traces the dust heating effectiveness
for a given starlight spectrum, and is normalized to that of the
mMMP radiation field: $\gamma_* > 1$ represents a harder radiation field than the mMMP field, and
$\gamma_* < 1$ is softer than mMMP.  Formally, $\gamma_*$ is
computed by:
\begin{equation}
\gamma_* \equiv \frac{\left[\left(\int d\nu \ u_{*,\nu} C_{\rm
      abs}\left(\nu\right) \right)/\int \ d\nu
    u_{*,\nu}\right]}{\left[ \left(\int d\nu \ u_{\rm mMMP,\nu} C_{\rm
      abs}\left(\nu\right)\right)/\int d\nu \ u_{\rm mMMP,\nu}\right]}
\end{equation}
\citet{draine21a} explicitly compute the difference in mid-IR features in response to different starlight spectra and intensities. We therefore need to explicitly compute this dimensionless parameter throughout our model galaxies, which we describe in \S~\ref{section:isrf}. 


The PAH emission spectrum is computed by \citeauthor{draine21a} for
individual PAH grain sizes, $14$ equally spaced log$U$ values from
log$U = [0,7]$, and a binary ionization state (neutral or ionized).
While this is a simplification, this parameterizes the PAH emission
spectrum in terms of physical input parameters that modern day galaxy evolution modeling techniques can track.  

In
the remainder of this section, we describe our methodology for
computing the dust content, grain size distribution, and ISRF from hydrodynamic simulations of galaxy evolution.

\subsection{Dust Content and Grain Size Distributions}
\label{section:dust_methods}
Our dust model is described in detail in Q. Li et al., (in prep), and
builds off of the framework developed by \citet{mckinnon18a} and \citet{li19a}.  We
describe this model, as well as updates to the \citet{mckinnon18a}
model here for completeness.

\subsubsection{Dust Formation}
The first major element in our model is to simulate the evolution of
the dust grain size distribution in galaxies.  To do this, we introduce a
new model for the formation, growth, and destruction of dust grains in
highly resolved galaxy evolution simulations.  In detail, we couple
this dust model with the {\sc smuggle} galaxy formation physics suite
and the {\sc arepo} hydrodynamics code (both are described in more detail
in \S~\ref{section:smuggle}), but in practice the methods that we
outline here are generalizable for any galaxy evolution model that
considers the evolution of the physical state of the ISM, as well as
stellar evolution processes.

Dust is produced through the condensation of metals that are returned
to the ISM by evolved stars.  Functionally in the models we produce simulated dust particles directly from simulated evolved star particles. We employ dust yields from
\citet{schneider14a} for Asymptotic Giant Branch (AGB) star dust production, and from
\citet{nozawa10a} for supernovae dust production.  The initial grain
size distribution for dust follows a lognormal size distribution:
\begin{equation}
    \frac{\partial n}{\partial a} = \frac{C}{a^p} \exp \left( \frac{\ln^2(a/a_{0})}{2 \sigma ^2} \right),
\end{equation}
where C is a normalization constant and $a_0 = 0.1
\micron$. $(p,\sigma)=(4,0.47)$ for dust produced by AGB stars and
$(p,\sigma)=(0,0.6)$ for SNII, following the work by
\citet{nozawa07a} and \citet{asano13a}.  This said, the results presented here are not strongly dependent on the initialized size distributions: due to interstellar processing of the dust grains, they quickly lose their memory of their initial size distributions (the relevant processes are described in more detail in \S~\ref{section:dust_evolution}).   We discretize the simulation dust particle sizes into $16$ size bins. This choice is arbitrary, though we find that this value results in converged size distributions in our simulations. 

In detail, we spawn new simulation dust particles from simulation star particles in a stochastic manner.   If a star particle of mass $M_*$ produces a dust mass $\Delta M_{\rm d}$ within a time step $\delta t$, we spawn a new dust particle of mass $M_d$ probabilistically if a randomly drawn number from a uniform distribution between $[0,1)$ is less than:
    \begin{equation}
    p_{\rm d} = \frac{M_*}{M_{\rm d}}\left[1-{\rm
          exp}\left(-\frac{\Delta M_{\rm d}}{M_*}\right)\right].
  \end{equation}
This stochastic production of dust particles mirrors the stochastic production star particles that has long been used in galaxy formation simulations~\citep[e.g.][]{springel03a} and ensures the total mass of dust spawned matches the integral of the dust production rate over long time periods.
We merge dust particles together if the mass of two neighboring particles is smaller than $0.1
    \times M_{*,\rm init}$, and split them if they grow to $10 \times
    M_{*,\rm init}$.

    The mass of dust produced by evolved stars follows the methodology of \citet{dwek98a}, with updated condensation efficiencies as described in \citet{li20a}. Following \citet{dwek98a}, the dust mass produced by AGB stars with a carbon-to-oxygen mass ratio C/O $>$ 1 is expressed as:
\begin{equation}
m_{i,d}^{\rm AGB}=
\begin{cases}
\delta_{\rm C}^{\rm AGB} \left(m_{C,{\rm ej}}^{\rm AGB} - 0.75 m_{O,{\rm ej}}^{\rm AGB}\right), & i\ =\ {\rm C}\\
0, & {\rm otherwise,}
\end{cases}
\end{equation}
where $\delta_i^{\rm AGB}$ is the condensation efficiency of element $i$ for AGB stars. The mass of dust produced 
by AGB stars with C/O $<$ 1 is expressed as
\begin{equation}
m_{i,d}^{\rm AGB}=
\begin{cases}
0, &  i\ =\ {\rm C}\\
16 \sum \limits_{i=\rm{Mg,Si,S,Ca,Fe}} \delta_i^{\rm AGB} m_{i, {\rm ej}}^{\rm AGB}/\mu, & i\ =\ {\rm O}\\
\delta_i^{\rm AGB} m_{i, {\rm ej}}^{\rm AGB}, & {\rm otherwise,}
\end{cases}
\label{eq:2}
\end{equation}
where $\mu$ is the mass of the species in amu.
 The mass of dust produced by Type II SNe is modeled as
\begin{equation}
m_{i,d}^{\rm SNII}=
\begin{cases}
\delta_{\rm C}^{\rm SNII} m_{{\rm C}, {\rm ej}}^{\rm SNII}, &  i\ =\ {\rm C}\\
16 \sum \limits_{i=\rm{Mg,Si,S,Ca,Fe}} \delta_i^{\rm SNII} m_{i, {\rm ej}}^{\rm SNII}/\mu, & i\ =\ {\rm O}\\
\delta_i^{\rm SNII} m_{i, {\rm ej}}^{\rm SNII}, & {\rm otherwise,}
\end{cases}
\label{eq:3}
\end{equation}
where $\delta_i^{\rm SNII}$ is the condensation efficiency of element
$i$ for SNII.  Here, we assume a fixed AGB condensation efficiency of
$\delta^{\rm AGB}_{i,\rm dust}=0.2$ \citep{ferrarotti06a} and
$\delta^{\rm SN II}_{i,\rm dust}=0.15$ \citep{bianchi07a}.  We assume two types of dust particles: silicates and carbonaceous.  For a given dust particle in our model, the total carbon mass in the dust
particle corresponds to the carbonaceous dust mass, and the remainder to
silicate.

  \subsubsection{Dust Evolution}
  \label{section:dust_evolution}

 Dust grains evolve from their initialized size distribution
 as they undergo growth from accretion, coagulation, thermal
 sputtering, shattering, and destruction in shocks and star forming regions.

Physical dust particles, with radius $a$, grow via the accretion of metals at a rate \citep{hirashita11a}:
 \begin{equation}
\left(\frac{da}{dt}\right) = \frac{a}{\tau_{\rm accr}}
 \end{equation}
 where the accretion timescale, $\tau_{\rm accr}$, is proportional to the size of the
 grain, and inversely proportional to the gas density, temperature,
 and metallicity:
 \begin{equation}
   \label{equation:growth}
\tau_{\rm accr} = \tau_{\rm ref}\left(\frac{a}{0.1 \mu {\rm m}}\right)\left(\frac{1000 \ {\rm cm}^{-3} \times  m_{\rm H} \times Z_\odot}{\rho_Z}\right)\left(\frac{10{\rm K}}{T_g}\right)^{\frac{1}{2}}\left(\frac{0.3}{S}\right)
 \end{equation}
where $\rho_Z$ is the metal density, $T_g$ is the gas temperature,
and $S$ is the sticking coefficient.  The growth timescale is limited
by the least abundant element required by the grain species following
\citet{choban22a}.  We adopt ($\tau_{\rm ref},Z_{\odot,{\rm SI}}$) = $\left(0.224 \ {\rm Gyr}, 7 \times 10^{-3}\right)$ for silicates \citep[assuming a composition of MgFeSiO$_4$ for silicates;][]{weingartner01a}, and ($\tau_{\rm ref},Z_{\odot,{\rm C}}$) = $(\left(0.175 \ {\rm Gyr}, 2.4 \times 10^{-2}\right)$ for carbonaceous grains.  We additionally adopt temperature dependent sticking coefficient following
\citet{zhukovska16a} which drops at higher temperatures (Q.Li et
al. in prep.).  This has the effect of significantly reducing the
growth rates in the warm and dense ISM that is heated by stellar
feedback (more on this in \S~\ref{section:smuggle}).

We include two forms of dust destruction: thermal sputtering and, in
star-forming regions, supernovae shocks.  For thermal
sputtering, grains can be eroded by hot electrons (which is especially
pertinent in the hot ISM and in hot halos):
\begin{equation}
\left(\frac{da}{dt}\right)_{\rm sp} = -\frac{a}{\tau_{\rm sp}}
\end{equation}
where the sputtering timescale, $\tau_{\rm sp}$ follows the analytic
approximation derived by \citet{tsai95a}, and is linearly proportional
to the grain radius, and inversely proportional to the gas density and
temperature.  Supernova shocks additionally destroy dust grains via thermal sputtering, where
the evolution in the grain size distribution follows the models of dust destruction in supernovae blastwaves by \citet{nozawa06a} and \citet{asano13a}.   The change rate of mass of grains in the $k$th size bin due to thermal sputtering is
\begin{multline}
\left( \frac{M_k}{t} \right)_{\rm de} =  \frac{\gamma M_{\rm s}}{M_{\rm g}} \times \\ \left\lbrace \sum_{i=1}^{N_{\rm bin}} \left[ N_i(t) \xi(a_k,a_i) \left( \frac{\pi \rho_{\rm gr} a^4}{3} \right) \right]_{a_k}^{a_{k+1}}- M_k(t) \right\rbrace .
\label{eq:11}
\end{multline}
Here, $M_{\rm g}$ is the gas mass, $\gamma$ is the rate of supernovae near the dust particle, $M_{\rm s}$ is the mass of neighboring gas swept by SN shocks (which is derived from \citealt{yamasawa11a}), and 
$\rho_{\rm gr}$ is the internal density of a dust grain with $\rho_{\rm gr} = 3.3$ and $2.2$~${\rm~g\;cm^{-3}}$ for silicate and carbonaceous grains, respectively.


Finally, we consider the impact of grain-grain collisions on the size
distribution of dust grains.  There are two important effects: dust
shattering, which results from high-speed encounters (and transforms
large grains into many small grains), and dust coagulation, which
results from low-speed encounters (and transforms small grains into
large grains).  Collision processes are mass conserving, but not
number conserving.  Following \citet{mckinnon18a} and \citet{li21a},
we model the transformation of grain sizes in collisional encounters by
the mass evolution of grain size bin $k$ by:
\begin{equation}
  \label{equation:shattering}
    \begin{aligned}
      \frac{{\rm d}M_k}{{\rm d}t} = -\frac{\pi \rho_{\rm d}}{M_{\rm d}} \Bigg( \sum_{k=0}^{N-1} v_{\rm rel}(a_i,a_k) m_i I^{i,k} - \\
      \frac{1}{2} \sum_{k=0}^{N-1} \sum_{j=0}^{N-1} v_{\rm rel} (a_k,a_j)  m^{k,j}_{\rm col}(i) I^{k,j}\Bigg)
  \end{aligned}
    \end{equation}
as long as the relative velocity between grains is greater than a
threshold velocity $v_{\rm rel} > v_{\rm thresh}$.  Here, the grain
sizes are denoted with $a$, $m_i$ is the mass of the grain in bin $i$,
and $m^{k,j}_{\rm col}\left(i\right)$ is the resulting mass entering
bin $i$ due to the collision between the grains in bins $k$ and $j$.
\citet{jones96a} suggest a threshold velocity of $v_{\rm thresh} =
2.7$ km s$^{-1}$ for silicates, and $1.2 $ km s$^{-1}$ for carbonaceous grains.
We implement a similar transfer of mass between size bins for dust
coagulation, though of course in this situation we only do so if
$v<v_{\rm thresh}$.  We follow \citet{hirashita09a} in employing a threshold velocity\footnote{There is clearly some freedom in computing
  these shattering rates: in particular, the choice of a threshold
  velocity where collisional processes transition from coagulation to
  shattering.  While a full exploration of the impact of the threshold
  velocities on the PAH population is outside the scope of what is
  computationally feasible here, we note that a similar implementation
  of dust collisional processes in the {\sc simba} simulation by
  \citet{li21a} of Milky Way-like galaxies in a cosmological
  simulation results in grain size distributions comparable to a
  \citet{mathis77a} ``MRN'' size distribution, and extinction laws
  comparable to the \citet{cardelli89a} Galactic constraints.  We have
  therefore adopted the same threshold velocity here, without any
  tuning.}  that is dependent on 
the grain sizes as well as the material properties of
the species, following their Equation~8.

\subsubsection{Converting between Aromatics and Aliphates}
\label{section:arom_ali}
In our model, all dust grains that are dominated by carbon are
considered carbonaceous; else, they are silicates.  Within the carbonaceous population,
dust grains are subdivided into aromatic hydrocarbons, and aliphatic
hydrocarbons.  We assume that the former represent PAHs.   We do not impose a size cut-off for PAHs, though note that larger PAHs are not effective emitters. We track both aromatics and aliphates, and
follow the methodology of \citet{hirashita20a} in converting between
the two\footnote{We note that more recent models by \citet{hensley22a} and references therein have advanced a picture of ``astrodust+PAHs'', where large grains are aggregated in their properties into astrodust, while nanoparticle aromatic carbonaceous grains are considered PAHs.  From the standpoint of the interface between our model and the \citet{draine21a} model, the treatment of the larger grains is less important than the PAHs themselves.  Once ultrasmall carbonaceous grains are aromatized in our model, we consider them PAHs, and they adopt the properties of the PAHs in the \citet{draine21a} and \citet{hensley22a} models.}.

In short, aromatization (i.e., the conversion from aliphates to
aromatics) is assumed to occur due to photoprocessing, and the removal
of hydrogen atoms from carbonaceous dust grains \citep[i.e.,
  dehydrogenation;][]{rau19a,hirashita20a}\footnote{While the removal
  of aliphatic side-groups from aromatic rings may also serve as a
  mechanism for aromatizing carbonaceous grains, we assume that this
  process is dominated by de-hydrogenization.  Future work will
  explore more detailed models for grain aromatization processes.}.
This is quantified by the change of the band gap energy, $E_g$, which
can be related to the fraction of hydrogen atoms in the grain, $X_{\rm
  H}$, via $E_g = 4.3 \times X_{\rm H}$ eV
\citep{tamor90a,hirashita20a}.  \citet{rau19a} and
\citet{hirashita20a} compute the timescale for full aromatization
(i.e., the time necessary to fully dehydrogenate the grain, from its
maximum assumed possible $X_{\rm H} = 0.6$ to its minimum assumed
value of $X_{\rm H} = 0.02$; \citealt{jones13a}) of:
\begin{equation}
  \label{equation:aromatization}
\frac{\tau_{\rm ar}^{\rm UV}}{\rm yr} = G_{\rm ISRF}^{-1} \left[3\left(\frac{a}{\mu m}\right)^{-2} + 6.6 \times 10^7 \left( \frac{a}{\mu m}\right)\right]
\end{equation}
where $G_{\rm ISRF}$ is the strength of the interstellar radiation
field, and $G_{\rm ISRF}=1$ corresponds to the Solar neighborhood\footnote{$G_{\rm ISRF}$ is related to $U$ in Equation~\ref{equation:intensity_parameter}.  We choose to employ two different variables here, however, in order to maintain consistency with other literature works.}
\citet{hirashita20a} compute this fitting formula for the
aromatization time via photo-processing by assuming a
\citet{mathis83a} radiation field shape.  This represents a minor
inconsistency in our modeling, as the local radiation field in
galaxies often departs from the \cite{mathis83a} fiducial shape, and
indeed a major aspect of \S~\ref{section:isrf} is to explicitly
account for these variations when computing the PAH emission spectrum.
Nevertheless, we proceed as an on-the-fly calculation of the exact
ISRF shape (which requires both knowledge of the stellar spectral
shapes, as well as the effects of the absorption and scattering of
photons by interstellar dust) is currently computationally intractable in
hydrodynamic galaxy evolution simulations.  We instead compute the
strength of the local interstellar radiation field by computing a
nearest neighbor search around each dust particle to get both the
location of nearby stars, as well as the dust column density between
that star and the dust particle of interest.  We compute a
mass-to-light ratio for these stars with {\sc fsps} \citep{conroy09a,conroy09b}, and assume a
\citet{weingartner01a} extinction law between the star particle and
the dust grain.  The incident radiation fields are then summed to
compute the local FUV flux.  We note that this extinction correction is not fully consistent with the grain size distribution and composition, though is necessary for computational feasibility.   

We assume that aromatic grains (i.e., those that we consider to produce the mid-IR PAH features) aliphatize via the accretion of free elements.
Using two-phase simulations, \citet{murga19a} and \citet{hirashita20a}
find a fitting function for the aliphatization rate in terms of the
grain size distribution:
\begin{equation}
\frac{\tau_{\rm al}}{{\rm yr}} = 1.6 \times 10^5 \left(\frac{a}{\mu {\rm m}}\right)
\end{equation}
\citet{hirashita20a} compute this fitting formula for dense gas, which
is defined as $n>300 $ cm$^{-3}$.  
We therefore implement a threshold
density of $n=300$ cm$^{-3}$ below which aromatic grains cannot aliphatize.  This serves
as a limiter for the conversion of aromatics to aliphates in diffuse
gas in our model. 

\subsection{Modeling the Interstellar Radiation Field}
\label{section:isrf}

With these pieces in hand, we now have a model for the evolution of
grain size distributions, and in particular, for the aromatic carbonaceous
component of the dusty ISM, in place.  The next stage is to compute
the emission properties of these aromatic hydrocarbons.  PAHs are
excited by ultraviolet and optical photons, and emit via vibrational transitions
as they cool.  Because of this, the shape of the ISRF matters significantly.  In order to model the ISRF, we
employ the publicly available {\sc powderday} dust radiative transfer
package\footnote{\url{github.com/dnarayanan/powderday}}
\citep{narayanan21a}, which employs {\sc yt}, {\sc fsps}, and {\sc
  hyperion} for grid generation, stellar population synthesis
calculations, and Monte Carlo radiative transfer respectively
\citep{turk11a,conroy09b,robitaille11a}.  We refer the reader to the
code paper \citep{narayanan21a}, as well as the documentation
site\footnote{\url{powderday.readthedocs.org}} for details, and here
describe the major attributes of {\sc powderday} that impact our
calculations, as well as updates since the \citet{narayanan21a} code
paper that we have implemented in order to enable this work.

We begin by computing the intrinsic stellar spectrum of all of the
stars in individual snapshots of our hydrodynamic simulations based on their ages and their metallicities as computed by the galaxy evolution model.  The stellar spectra of these stars are computed using {\sc fsps} \citep{conroy09b,conroy10a,conroy10b}.  We assume the {\sc mist} stellar isochrones
\citep{choi16a}, and a \citet{kroupa02a} stellar initial mass
function. This light is emitted
in an isotropic manner in a Monte Carlo fashion through the
interstellar medium of the galaxy.  As we will describe in
\S~\ref{section:smuggle} (though in principle these methods are
generally applicable to a range of types of galaxy simulations), we
conduct our hydrodynamic simulations with the {\sc arepo} code on a
Voronoi mesh.  The radiation therefore propagates through this Voronoi
mesh, and encounters the dust content and size distribution as computed in
\S~\ref{section:dust_methods}.

Informed by the grain size distribution in every cell, we compute the
local extinction law on a cell-by-cell basis following the methods
outlined in \citet{li21a}.  In detail, the optical depth at a wavelength $\lambda$ can be computed in terms of an extinction efficiency:
\begin{equation}
\tau\left(a,\lambda\right) da = \int_{\rm LOS} \pi a^2 Q_{\rm ext}\left(a,\lambda\right)n_{\rm d}\left(r,a\right)da \ ds
\end{equation}
where $Q_{\rm ext}$ is the extinction efficiency, and $n_{\rm
  d}\left(r,a\right)$ is the number density of grains with sizes
$[a,a+da]$ (we do not assume any subresolution clumping: the density is the number of grains divided by the cell volume).  We assume the efficiencies $Q_{\rm ext}$ as computed by
\citet{draine84a} and \citet{laor93a} for silicates and carbonaceous grains,
respectively, though in future versions of this code we will implement
the newly developed ``astrodust'' size-dependent extinction properties
of dust grains \citep{hensley22a}.  
The wavelength dependent extinction is:
\begin{equation}
  \begin{aligned}
    A\left(\lambda\right) = 2.5 {\rm ln} \int_{\rm a_{\rm min}}^{a_{\rm max}} \tau \left(a,\lambda\right)da  = \\
    2.5\,{\rm ln} \int_{\rm a_{\rm min}}^{\rm a_{\rm max}}da \pi a^2 Q_{\rm ext}\left(a,\lambda\right) \int_{\rm LOS} n_d\left(r,a\right) ds
  \end{aligned}
  \end{equation}

Having computed the local extinction law in every cell in the model
galaxy, we proceed with the Monte Carlo dust radiative transfer.  The
direction and frequency of photons are randomly drawn, and the photons
are propagated until they either escape the grid, or reach a randomly
drawn optical depth drawn from an exponential distribution.  Formally,
$\tau = -{\rm ln}\epsilon$ where $\epsilon = [0,1)$.  Photons can be
  either scattered or absorbed at this point, depending on the dust
  albedo.  This procedure is iterated upon until the equilibrium dust
  temperature is converged.  Upon convergence, we have the mean
  interstellar radiation field in every cell in the galaxy at
  wavelengths longer than $\lambda > 912 \angstrom$.

 The \citet{draine21a} model emission spectra for PAHs are
 pre-computed per dust grain size for $14$ different incident
 radiation fields and $15$ log$U$ intensities.  Because the radiation
 fields in any individual cell are not likely to be well-described by
 any of the $14$ individual pre-computed stellar SEDs employed in the
 \citet{draine21a} model, we employ a non-negative least squares
 decomposition.  This is akin to treating the \citet{draine21a}
 input SEDs as basis functions, and determining the coefficients in
 front of these basis functions that allow for a linear combination
 into a cells individual radiation field.

 At this point, for every cell we know how the local ISRF can be decomposed
 into the \citet{draine21a} basis SEDs, as
 well as the individual grain size distribution.  The final step in
 this model is to compute the ionization state of the PAHs in every
 cell.  We follow \citet{draine21a} and \citet{hensley22a}, and utilize the analytic
 relationship between grain size and ionization fraction:
 \begin{equation}
 \label{equation:ionization}
f_{\rm ion}\left(a\right) = 1 - \frac{1}{1+a/10 \angstrom}
 \end{equation}
 In reality the grain ionization state will depend on local conditions, such as the density and incident radiation field.  We discuss the model prospects for implementing a more sophisticated model for grain ionization in \S~\ref{section:discussion}.  
 
 The PAH emission spectrum in each individual cell can now be computed as a
 summation of the PAH emission spectra per grain size, at a given
 log$\left(U\right)$, ionization state, and fractional contribution
 per basis SED incident radiation field.  These PAH emission spectra
 are then added as source terms, and the dust radiative transfer is
 re-iterated upon, this time including the PAHs.  The purpose of this
 second iteration is to capture any potential attenuation of PAHs in
 extremely dense environments.  Indeed at least some simulations have
 suggested extreme optical depths in dusty galaxies such as model
 submillimeter galaxies \citep{lovell22a}.   

 Finally, the aggregate SED from model galaxies are computed via ray
 tracing.  Here, the source function $S_\nu$ is computed at each
 position in the grid, and then we integrate the equation of radiative
 transfer along each line of sight at multiple viewing angles
 surrounding the model galaxy.

\subsection{Galaxy Evolution Simulations}
\label{section:smuggle}
\begin{figure*}
  \includegraphics[width=\textwidth]{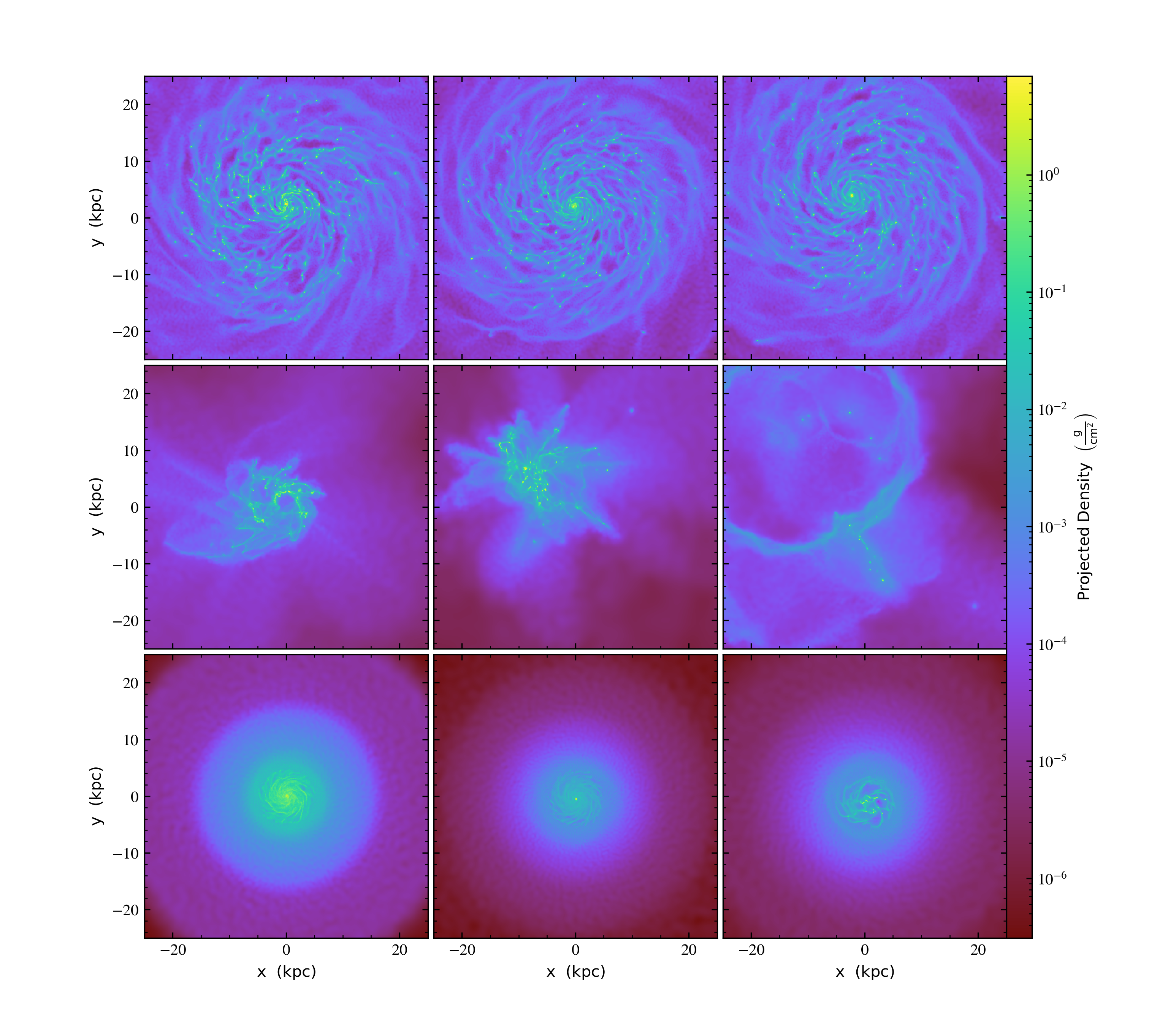}
  \caption{{\bf Gas surface density evolution for $3$ galaxy models
      presented in this work: Milky Way (MW), Sbc and Dwarf} (from top to bottom). All
    image panels are the same scale so that the relative differences
    in the galaxy sizes are evident.  The columns (left to right) are at $1$, $2.5$ and $4.5$ Gyr. Models MW and Dwarf are designed
    to serve as analogs to the Milky Way, and a low-metallicity Dwarf
    galaxy, while model Sbc is designed to experience relatively high
    gas surface densities, and hence large feedback events.  The
    models here intentionally span a diverse range of physical
    conditions.  
   \label{figure:proj}}
\end{figure*}
We have implemented the aforementioned model into the {\sc arepo}
hydrodynamic code base \citep{springel10a,weinberger20a}, with the
Stars and MUltiphase Gas in GaLaxiEs ({\sc smuggle}) galaxy formation
physics suite enabled \citep{marinacci19a}.  Here, we describe the
relevant details, though note that in principle the methods for the emission properties of PAHs described
thus far are agnostic to the actual galaxy formation code and
implemented physics.

Primordial cooling occurs via two-body collisional processes,
recombination, and free-free emission \citep{katz96a}, as well as
Compton cooling off of CMB photons.  Metal-enriched gas undergoes
metal line cooling, whose rates are computed as a function of
temperature and density based on {\sc cloudy} photoionization
calculations \citep{ferland13a}, as described in
\citet{vogelsberger13a}. Low-temperature cooling (T$\sim 10-10^4$ K)
can occur via metal line, fine-structure, and molecular cooling
processes via a fit to the \citet{hopkins18a} {\sc cloudy} cooling
tables as presented in \citet{marinacci19a}.  Here, gas at densities
$n>10^{-3}$ cm$^{-3}$ can self-shield, following the
\citet{rahmati13a} parameterization.  The self-shielding processes are
redshift-dependent: because the simulations presented here are
idealized disk galaxies, we adopt the $z=0$ scalings from Table A1 of
\citet{rahmati13a}.  At the same time, gas can be heated both by
cosmic rays, as well as photoelectric processes.  Cosmic ray heating
follows the density-dependent prescription of \citet{guo08a}, while
photoelectric heating follows the density, metallicity, and
temperature-dependent rates derived by \citet{wolfire03a}.  Details
for the implementation of both heating rates are given in Equations
$3$--$6$ of \citet{marinacci19a}.

Star formation occurs in gravitationally bound gas \citep{hopkins13e}
above a specific density threshold.  We set this threshold for our
simulations to $n_{\rm thresh} = 1000$ cm$^{-3}$.  Star formation in
this gas occurs probabilistically \citep{springel03a} following a
volumetric \citet{kennicutt98a} relation such that:
\begin{equation}
\dot{M}_* = \epsilon \frac{M_{\rm gas}}{t_{\rm ff}}
\end{equation}
where $\epsilon$ is the star formation efficiency, $\dot{M_*}$ is the star formation rate, $M_{\rm gas}$ is the gas mass, and $t_{\rm ff}$ is the gas free fall time:
\begin{equation}
t_{\rm ff} = \sqrt{\frac{3\pi}{32 G \rho_{\rm gas}}}
\end{equation}
We set the star formation efficiency factor to $\epsilon=1$:
\citet{hopkins18a} demonstrate that the effective star formation
efficiency for explicit feedback models such as this are relatively
insensitive to this choice \citep[see the review by][]{vogelsberger20a}.  We limit star formation to occur
exclusively in molecular gas and compute the molecular gas fraction
via the \citet{krumholz08a,krumholz09a,krumholz09b} prescription
linking the H$_2$ fraction to the local gas surface density and
metallicity.  

Once formed, stars return energy to the ISM.  The fraction of Type I
and II supernovae are computed from each star particle by assuming a
\citet{chabrier01a} stellar initial mass function, though the former
also includes a delay time distribution in deriving the number of Ia
events \citep{vogelsberger13a}.  The details for mass loss rates, and
energy and momentum coupling to the ISM are detailed in
\citet{marinacci19a}.  Similarly, young stars impart feedback into the
nearby ISM via radiation.  Here, feedback is included from young star
photoionization, radiation pressure, and OB and AGB stellar winds.
Taken together, these feedback mechanisms act to both regulate the
star formation rates in our model galaxies, as well as impact ISM
densities, temperatures, and velocity dispersions that can determine
critical dust processes such as growth rates and grain velocities.

\begin{table*}
        \centering
        \caption{Model Simulations Used in this Paper }
        \label{table:models}
        \begin{tabular}{lccccccc}
                \hline
                Name & Galaxy Type & M$_{\rm halo}$ & $c_{\rm halo}$ & $M_{\rm disk}$ & $M_{\rm gas}$ & $M_{\rm Bulge}$ & $m_{\rm gas}$\\
                 & & $M_\odot$ & & $M_\odot$ & $M_\odot$ & $M_\odot$ & $M_\odot$\\
                \hline
                \hline
                MW & Milky  Way Analog & $1.5 \times 10^{12}$ & 12 & $4.7 \times 10^{10}$ & $9 \times 10^{9}$ & $1.5 \times 10^{10}$ & $3220$\\
                Sbc & Gas-Rich Starburst & $1.5 \times 10^{11}$ & 11 & $4 \times 10^{9}$ & $5 \times 10^{9}$ & $10^9$ & $1.8 \times 10^4$\\
                Dwarf & Low Metal Dwarf & $2 \times 10^{10}$ & 15 & $1.3 \times 10^{8}$ & $7.5 \times 10^8$ & $10^7$ & $2500$\\
                \hline
        \end{tabular}
\end{table*}

\section{Evolution of Galaxy and Dust Physical Properties}
\label{section:galaxy_evolution}
\subsection{Galaxy Model Overview}
In this paper, we examine the physical and emission properties of PAHs
from $3$ idealized galaxy evolution simulations.  These simulations,
while initialized to broadly resemble properties of different galaxy
types observed in the local Universe, are not by any means intended to
serve as specific analogs to any individual galaxy.  Instead, our goal
is to simulate a diverse range of physical conditions in order to
build a physical foundation for what drives variations in dust grain
properties in galaxies, and how that impacts their PAH masses and
luminosities.  Future work will include bona fide cosmological
simulations with realistic evolutionary histories, and detailed comparisons to large samples of galaxies in the JWST era.

\subsection{Initial Conditions and Galaxy Models}
We set up equilibrium initial conditions for three idealized galaxy
models (described shortly) following the technique first described by
\citet{springel05a}.  Galaxies are initialized with gaseous and
stellar disks, an old star bulge, and embedded in a live dark matter
halo with a \citet{hernquist90a} density profile.  The disk components
are exponential radially, though the stellar disk follows a sech$^2$
vertical profile.  The halo concentration and virial radius for a halo
of a given mass are motivated by $N$-body cosmological models
following \citet{bullock01a} and \citet{robertson06a}. We follow
\citet{hopkins11b}, and initialize $3$ model galaxies: a Milky Way
analog (MW), a Dwarf galaxy, and a starburst disk (Sbc).  These vary
not only in mass, but initial stellar and gas fractions.  We summarize
the initial condition parameters in Table~\ref{table:models}.

\begin{figure*}
  \includegraphics[]{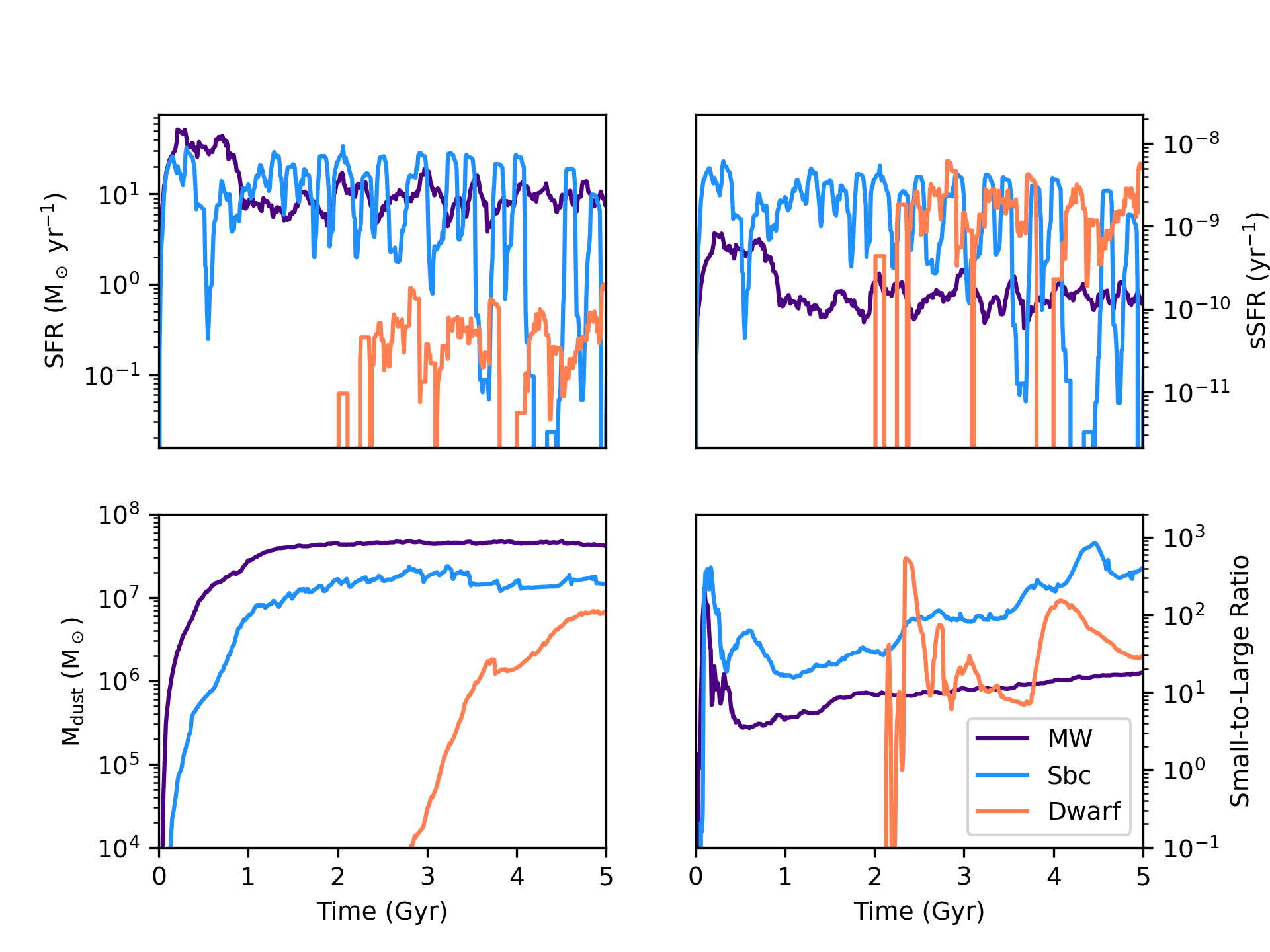}
  \caption{{\bf Evolution of physical properties for $3$ galaxy models
      presented in this work: Milky Way (MW), Sbc and Dwarf.}.
    Clockwise from top left, we show the: star formation rate
    evolution, specific star formation rate (sSFR$\equiv$SFR/$M_*$),
    ratio of small grains ($a<13 \angstrom$) to large dust grains
    ($a>13 \angstrom$), and dust mass.  The physical origin for these
    trends is discussed in
    \S~\ref{section:physical_properties}. \label{figure:physical_properties}}
\end{figure*}
\subsection{Galaxy Physical Property Evolution}
\label{section:physical_properties}

We first orient the reader to the evolution of the galaxy physical
properties by showing their gaseous morphologies at various
evolutionary points between t=[$0,4$] Gyr (Figure~\ref{figure:proj}),
as well as the evolution of four key physical properties in
Figure~\ref{figure:physical_properties}.  In particular, we present
(clockwise from the top left): the SFR, specific SFR (sSFR$\equiv$SFR/$M_*$), dust mass
and mass ratio of small grains to large grains.  We discuss these
panels in turn.

The galaxy SFRs are bursty, though oscillate around relatively steady
values.  This is an effect of the feedback model, which results in
relatively self-regulated star formation histories that proceed in a
quasi-equilibrium state \citep{marinacci19a}, and has been extensively
documented in other explicit feedback models such as {\sc fire}
\citep{hopkins11b,hopkins18a,gurvich22a}.

In the top right panel of Figure~\ref{figure:physical_properties}, we
show the galaxy specific star formation rates (SFR/$M_*$) for our
three idealized galaxy models.  While the total star formation rates
for the Sbc and MW are significantly greater than the Dwarf galaxy
model, we now see that the sSFRs for the Sbc and Dwarf are a factor
$\sim 50$ greater than the MW model.  This is due to our choice of
parameters in the initial conditions (Table~\ref{table:models}): specifically, both the Dwarf and Sbc models are initialized with a
significantly higher gas fraction than the MW, and at least as high of
a halo concentration.  This results in high gas densities, and
subsequently increased star formation rates per unit stellar mass in
these lower mass systems.  As we will discuss throughout the remainder
of this paper, this will have an impact on the grain size
distributions and PAH properties in these galaxies.

For each model galaxy, the dust masses grow rapidly early on, and
eventually stabilize (bottom left of
Figure~\ref{figure:physical_properties}).  The growth of dust masses
owes first to formation in evolved stars, but is then dominated by
growth via metal accretion \citep{li19a,whitaker21a}.  At the same
time, this is balanced by the destruction of dust via thermal
sputtering.  Because the implemented feedback models regulates star
formation in a quasi-equilibrium state, the growth rates remain
relatively constant owing to steady metal injection into the
ISM. Similarly, the rates of dust destruction also remain relatively
constant owing to a lack of rapid variations in the galaxy physical
properties \citep{marinacci19a}.  It is important to note that while
this is the case for idealized galaxies in a self-regulated model,
this sort of quasi-equilibrium evolution in physical properties may
not necessarily apply to galaxies evolving in a cosmological context.

Finally, in the bottom right of
Figure~\ref{figure:physical_properties}, we show the evolution of the
ratio of small grains to large grains for each of our models. This
ratio is defined here as the ratio of mass of grains smaller than $13 \angstrom$
(the size for a PAH dust grain with $<1000$ carbon atoms) to mass of
grains greater than $13 \angstrom$, though the trends are robust against
reasonable changes to this fiducial value.  The size evolution of dust
grains is primarily dictated in our model by the ratio of shattering
to coagulation.  Shattering processes result in large grains
transitioning to smaller grains.  Models Sbc and Dwarf have relatively
large specific star formation rates compared to model MW, and hence
inject more energy per unit mass into the ISM than model MW.
The increased velocity in the ISM drives up shattering rates
(Equation~\ref{equation:shattering}), and increases the small to large
ratio.  It is these small dust grains that have the potential to emit
in the mid-IR, and will be of interest in the remainder of this paper.



\section{Demonstration of Methods: PAH Spectra and Images}
\label{section:demonstration_of_methods}
\begin{figure}
      \includegraphics[width=\columnwidth]{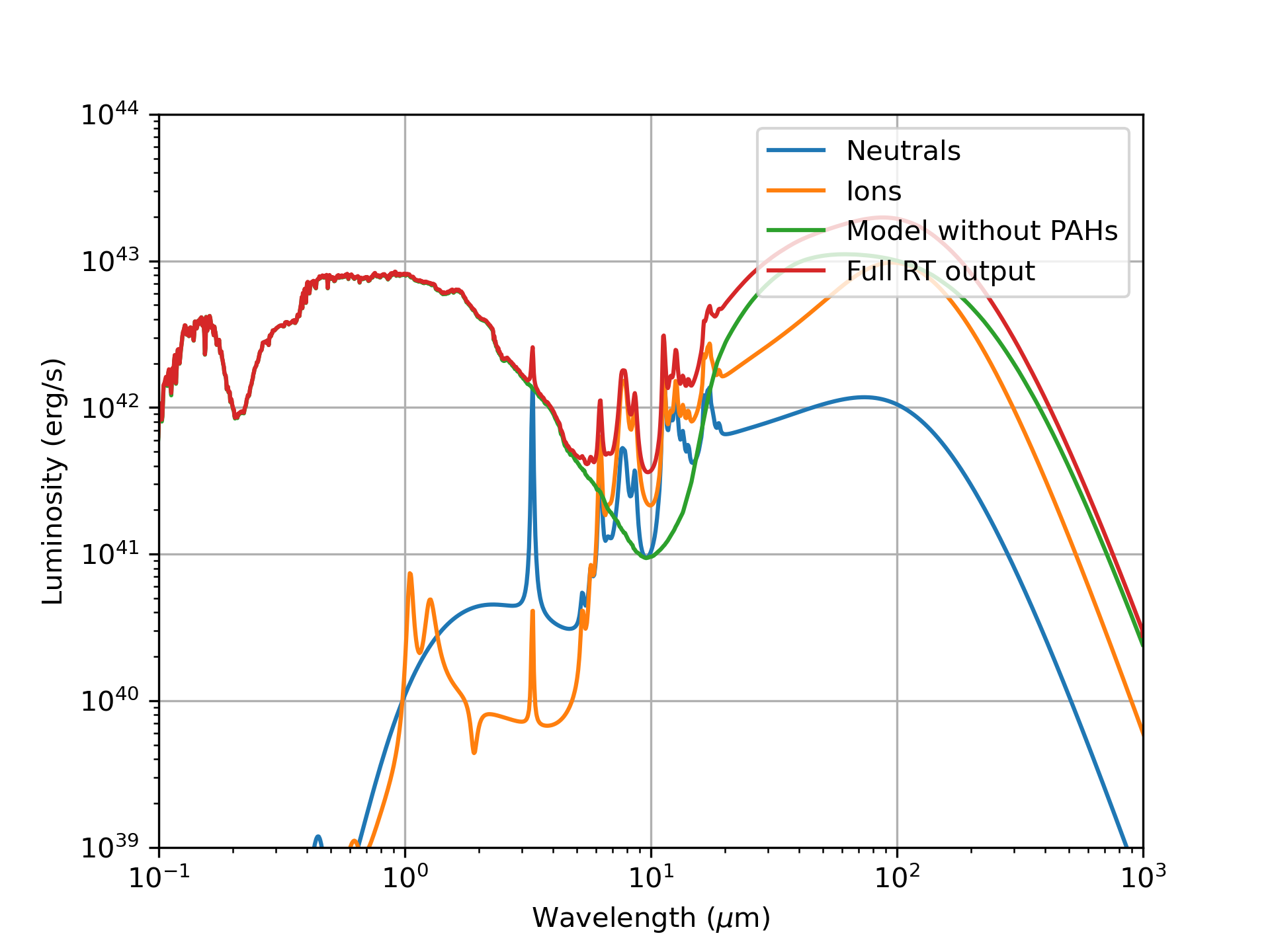}
      \includegraphics[width=\columnwidth]{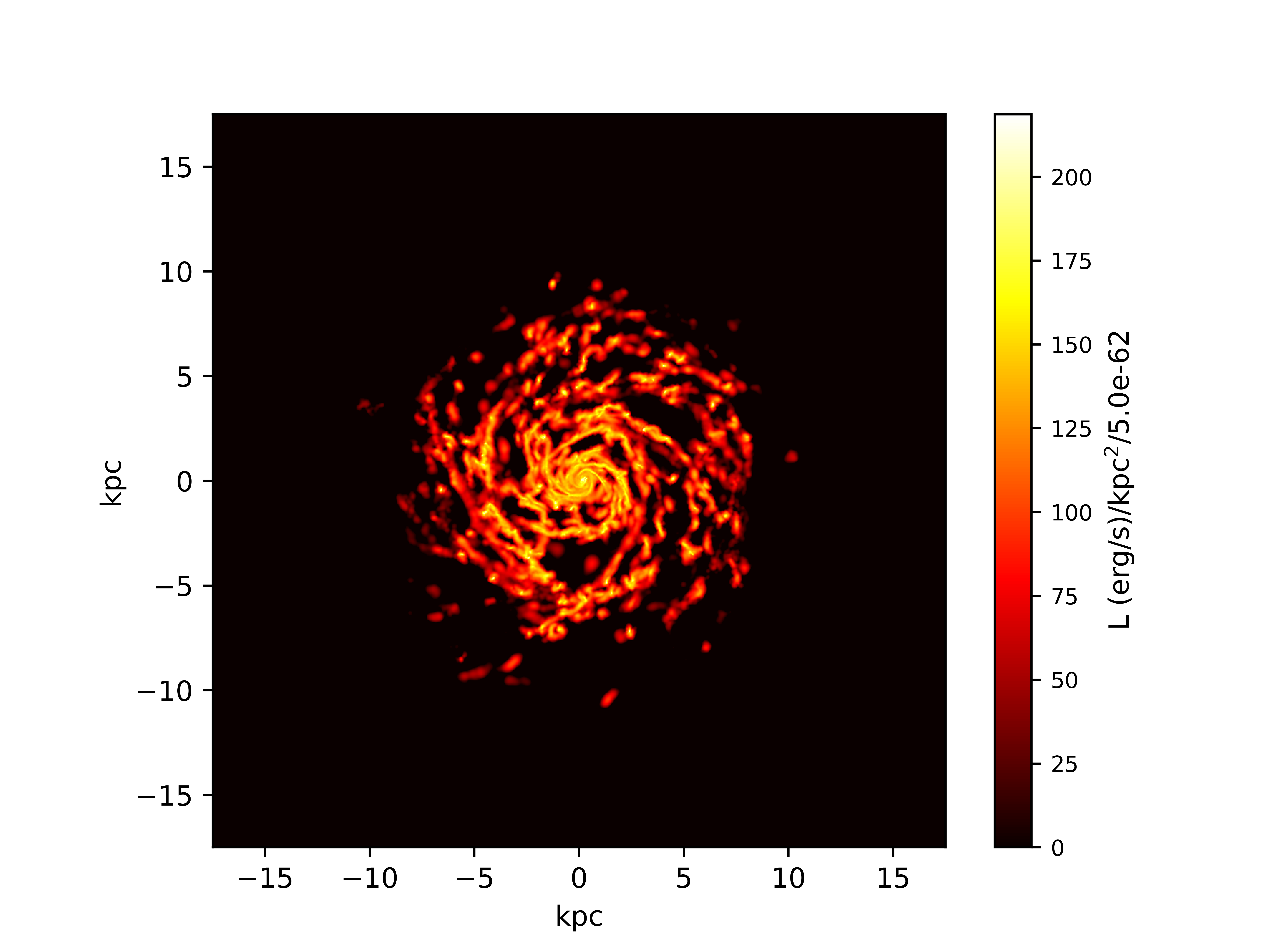}
      \caption{{\bf Demonstration of methodology: PAH SED and Surface
          Brightness Image of model MW.}  {\it Top:}
        Ultraviolet-millimeter wave SED.  We show the full radiative
        transfer output in red (i.e., what would be observed), as well
        as the model without PAHs (green), and the individual
        contribution of ionized PAHs (orange) and neutral PAHs (blue).
        {\it Bottom:} PAH surface brightness image (integrated over
        all bands from $3.3-20$ \micron) of the same model, in face-on
        view.\label{figure:sed_decomposed}}
    \end{figure}

Prior to studying our main results -- the origin of PAH masses and
luminosities in galaxies -- we first demonstrate the capabilities of
this new model by presenting a model galaxy SED and integrated PAH
surface brightness image in Figure~\ref{figure:sed_decomposed}.  In
the top panel of Figure~\ref{figure:sed_decomposed}, we show the SED
at an arbitrary time stamp for model MW, decomposed into the
following: the full observed SED, the SED without including our model
for PAH emission, and the contribution of ionized and neutral PAHs to
the mid-IR spectrum. 

In the bottom panel of Figure~\ref{figure:sed_decomposed}, we show the
PAH surface brightness, integrated over all bands from [$3.3-20$]
\micron \ for the same galaxy.  The variation of PAH surface brightnesses
across the galaxy owes to varying radiation field strengths, hardnesses,
and dust grain size distributions.  While a detailed analysis of the
drivers of resolved PAH brightness variations within an individual
galaxy model are outside the scope of this work, we study the impact
of these processes on the global \lpah \ from our models in \S~\ref{section:luminosity}.

    \section{PAH Masses and Luminosities in Galaxies}
    \label{section:masses_and_luminosities}
    \subsection{\texorpdfstring{PAH Masses and \qpah \ Fractions}{PAH Masses and $q_{\rm PAH}$ Fractions}}
    \label{section:qpah}


    \begin{figure*}
      \includegraphics{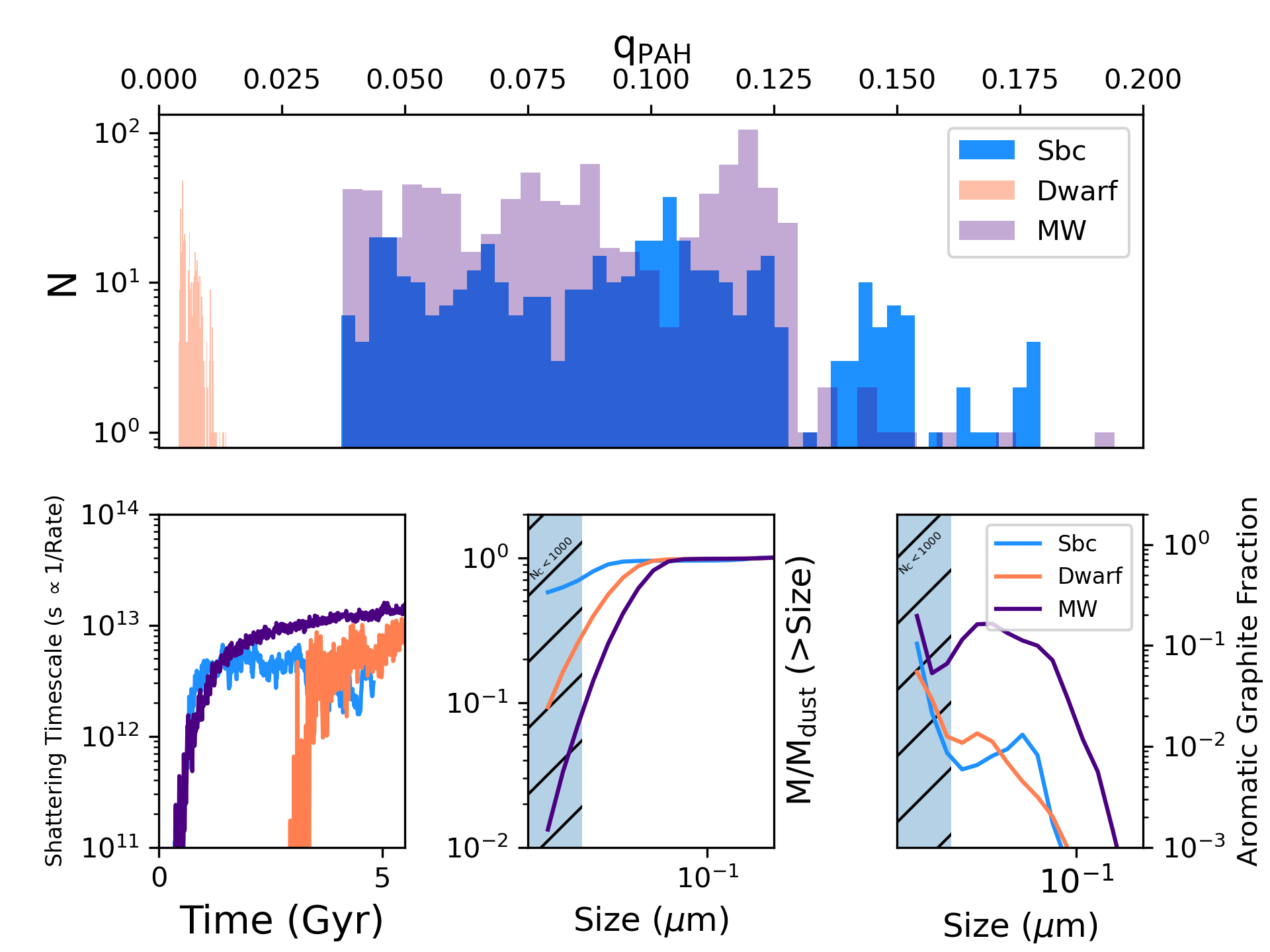}
      \caption{{\bf Distribution of $q_{\rm PAH}$ PAH mass fractions
          for model galaxies, and their relation to grain physics.}
        The model dwarf galaxy has relatively low PAH mass fractions
        ($\sim 1$--$2\%$), while the MW peaks at $\sim 5$--$10 \%$.  Both of
        these are in reasonable agreement with observational
        constraints from the Magellanic Clouds \citep{chastenet19a},
        as well as nearby disk galaxies \citep{draine07b}.  The bottom
        panels show (from left to right) the shattering timescale for
        each galaxy (which is inversely proportional to the shattering
        rate), the cumulative grain size distribution at a fixed time
        (normalized by the total dust mass), and the aromatic
        grain fraction at a fixed time.  The combination of small
        grain fractions (which are driven by the shattering rates)
        with the aromatic fractions set the individual \qpah \ PAH
        mass fractions in our models. The relevant discussion for this
        Figure is in \S~\ref{section:qpah}.\label{figure:qpah}}
    \end{figure*}

       \begin{figure*}
      \includegraphics{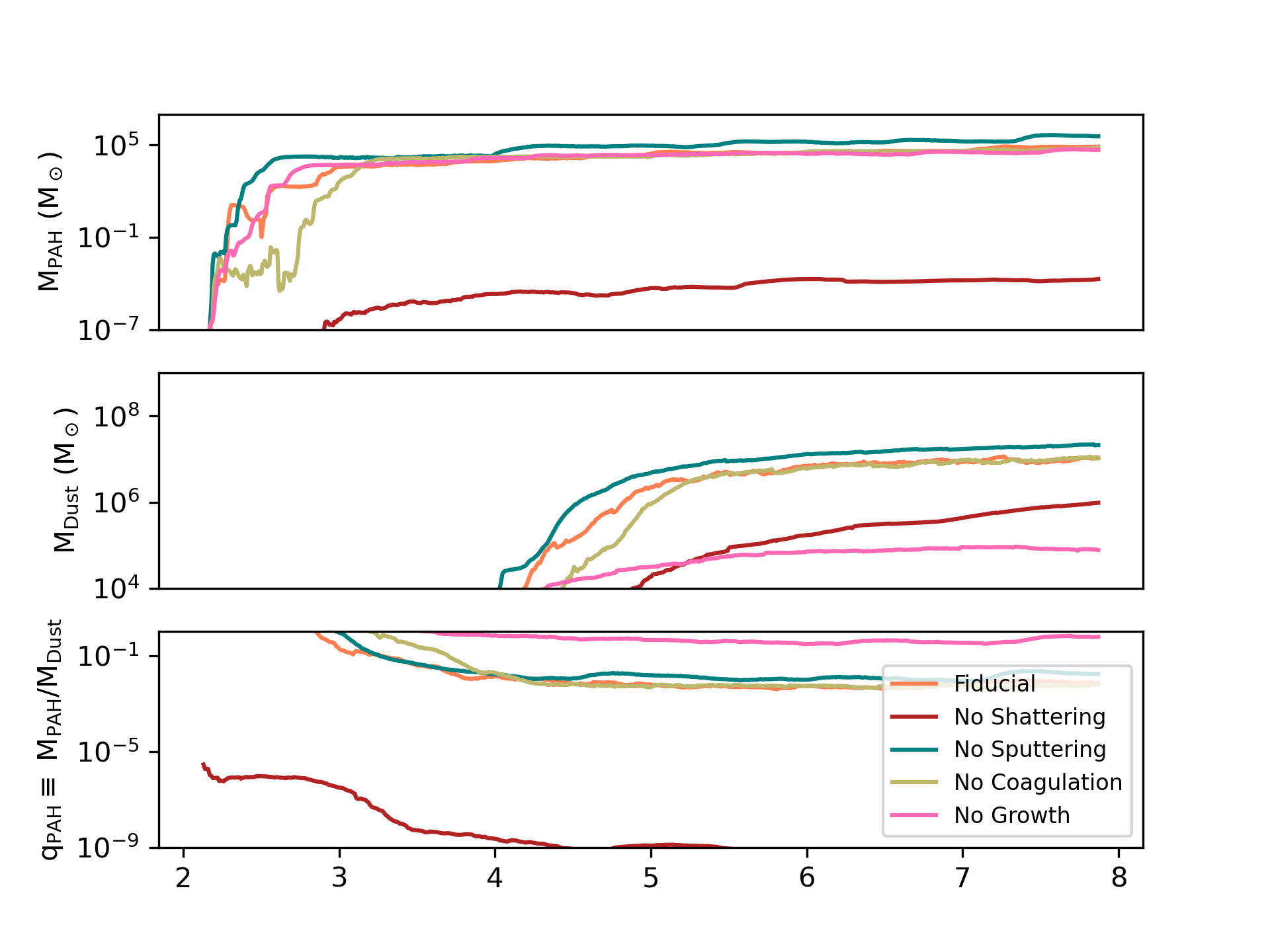}
      \caption{{\bf Sensitivity of dust and PAH masses to underlying
          dust physical processes.}  We conduct numerical experiments
        with our model dwarf galaxy by turning off various aspects of
        the dust model.  We find that grain-grain shattering is the
        most important element in creating PAHs (as this is the most
        efficient means of creating ultra small grains), while dust
        growth via metal accretion in the ISM is the most important process for setting overall
        dust masses. The relevant discussion for this Figure is in
        \S~\ref{section:qpah}. \label{figure:qpah_tuner}}
    \end{figure*}

We begin our analysis by studying the build up of PAH mass in our
model idealized galaxy evolution simulations.  We focus in particular on the {\it fraction} of dust mass that is in the form of PAHs.  This is traditionally defined in the literature as:
\begin{equation}
q_{\rm PAH} \equiv \frac{M_{\rm PAH}}{M_{\rm dust}}
\end{equation}
and is usually defined for PAHs containing $<10^3$ carbon atoms \citep{draine07b}.   \citet{hensley22a} relate the number of carbon atoms to the PAH size via:
\begin{equation}
  N_{\rm C}^{\rm PAH} = 1 + \left[417 \times \left(\frac{a}{10 \angstrom}\right)^3\right]
\end{equation}
which we adopt here. In the top panel of Figure~\ref{figure:qpah}, we
show histograms of $q_{\rm PAH}$ for our three model galaxies: MW,
Dwarf and Sbc.  There is generally a trend in which the Dwarf has the
lowest \qpah \ values ($\sim 1$--$2\%$), followed by the MW ($\sim
1$--$10\%$), with the Sbc starburst model displaying an extremely broad
range, including relatively large values as compared to local galaxies
(up to $\sim 15\%$).  For reference, observed constraints of \qpah
\ for the SMC and LMC are in the $\sim 1-2\%$ range
\citep{chastenet19a}, while local disk galaxies from the SINGs survey
typically exhibit PAH mass fractions $\sim 1-5\%$ \citep{draine07b}.
In what follows, we dissect the PAH mass fractions seen in our model
galaxies.

In short, the PAH fraction in our models is dictated by two physical
effects: the dust grain size distribution (and in particular, having
significant mass in the lowest size bins), and the aromatization of carbonaceous grains. 
We discuss these processes in turn, and highlight key differences
 between the idealized galaxy models presented here. 

We first begin with an examination of the
relevant dust processes in driving PAH masses with a controlled
numerical experiment.  In Figure~\ref{figure:qpah_tuner}, we show the
time evolution of the dust masses, PAH masses, and ratio of the two
($q_{\rm PAH}$) for our fiducial Dwarf model\footnote{We chose this model to explore physical variants of the dust model as the run time is the shortest.}, as well as $4$ model variants:
runs with shattering turned off, sputtering turned off, coagulation
turned off, and growth turned off.  The build up of total dust masses
are dominated by dust growth.  Turning off either dust growth
or shattering severely suppresses the total dust mass in the
model galaxies (the latter process is due to the need for shattering
to create small grains, and the dust growth timescales' linear
dependence on grain sizes; c.f. Equation~\ref{equation:growth}).  At
the same time, the total PAH masses are dominated almost entirely by
grain-grain shattering processes.  In principle, PAHs can form in our model from either the growth (via metal accretion) of the smallest grains as they are injected into the ISM from evolved stars, or from the shattering of larger grains.  In our model, shattering is by far the most efficient
means for transferring dust mass from larger sizes to small grains.  As
a result, the model \qpah \ values are relatively insensitive to the
underlying dust physics aside from two key processes: dust growth in
order to set the total dust masses in galaxies, and grain-grain
shattering in order to create ultra small grains in the ISM.

We are now in a position to understand the model \qpah \ distributions
in the top panel of Figure~\ref{figure:qpah}.  Recalling the galaxy
physical properties presented in
Figure~\ref{figure:physical_properties}, our model starburst galaxy
(Sbc) has the highest star formation rate per unit galaxy mass,
followed by the Dwarf model, with the MW analog exhibiting the lowest
sSFR.  The ISM velocity dispersions, and hence dust shattering rates,
correlate with these sSFRs.  In the bottom left panel of
Figure~\ref{figure:qpah}, we show the shattering timescale (which is
inversely related to the shattering rate) of the three model galaxies.
The model Sbc has the shortest shattering timescales, followed by the
Dwarf, with the MW having the longest timescales.  This has a direct
impact on the modeled grain size distributions. In the bottom-middle
panel of Figure~\ref{figure:qpah}, we show the cumulative dust grain
size distribution at a fixed time ($\sim 5$ Gyr; normalized by the
total dust mass) for each of our models, and demarcate the region that
would be considered PAHs.  Following the shattering rates, the Sbc
model has the largest fraction of its dust grains in the smallest size
bins, followed by the dwarf model, followed by the MW.

Taken at face value, this seems at odds with the findings in the top
panel of Figure~\ref{figure:qpah}, which shows a larger average \qpah
\ for the MW model than the Dwarf.  
The key lies in the aromatization of these smallest grains.  In the
bottom right panel of Figure~\ref{figure:qpah}, we show the cumulative
aromatic fraction for the three model galaxies in the same
size bins.  Here, the total SFR is the relevant quantity: the
aromatization rate is directly dependent on the incident FUV flux,
which increases with the local SFR density for a given dust particle.
This drives a larger fraction of the smallest grains for the MW model
to convert to PAHs than the Dwarf model, resulting in higher \qpah
\ fractional PAH ratios.  It is worth noting, however, that additional destruction mechanisms that are not implemented (such as photodestruction) may temper these trends: we discuss this further in \S~\ref{section:discussion}.






  \subsection{Drivers of PAH Luminosities}
  \label{section:luminosity}

  \begin{figure*}
      \includegraphics{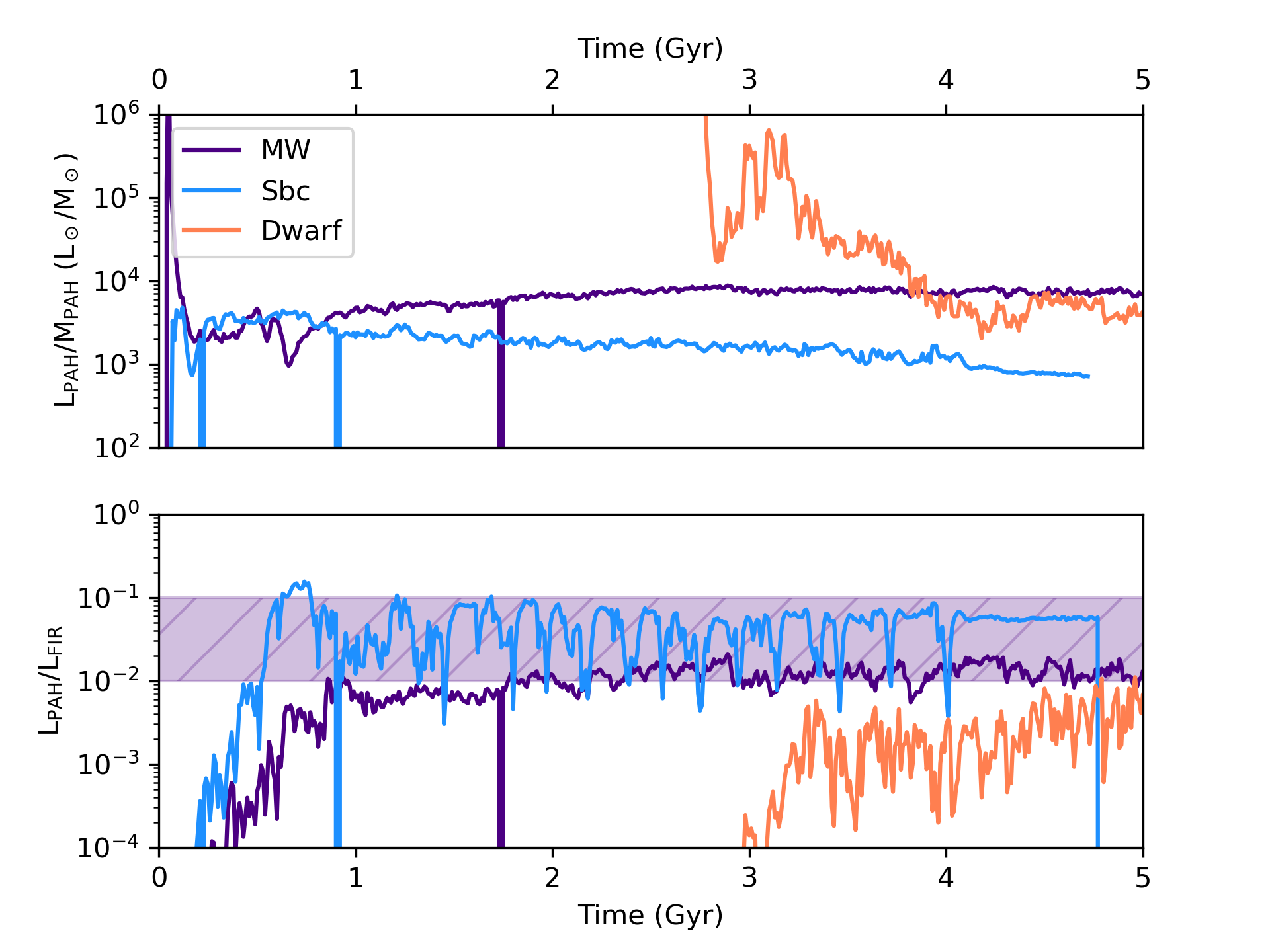}
      \caption{{\bf Evolution of PAH luminosity (integrated over all
          bands) for each galaxy model, normalized by their total PAH
          masses (top) as well as the total FIR luminosity (bottom)}.
        The purple shaded region in the bottom panel shows the range
        of observed $L_{\rm PAH}$ - $L_{\rm FIR}$ ratios in the local Universe
        \citep{smith07a}.  The PAH luminosities are a consequence of
        varying interstellar UV luminosities and grain size
        distributions.  Of note, the models presented here naturally
        fall into the observed range of $L_{\rm PAH}$ - $L_{\rm FIR}$ ratios,
        which is a consequence of \qpah \ fractions $\sim 1-10\%$, as
        well as galaxy star/dust geometries comparable to those in the local
        Universe. \label{figure:lpah}}
    \end{figure*}

      \begin{figure*}
      \includegraphics{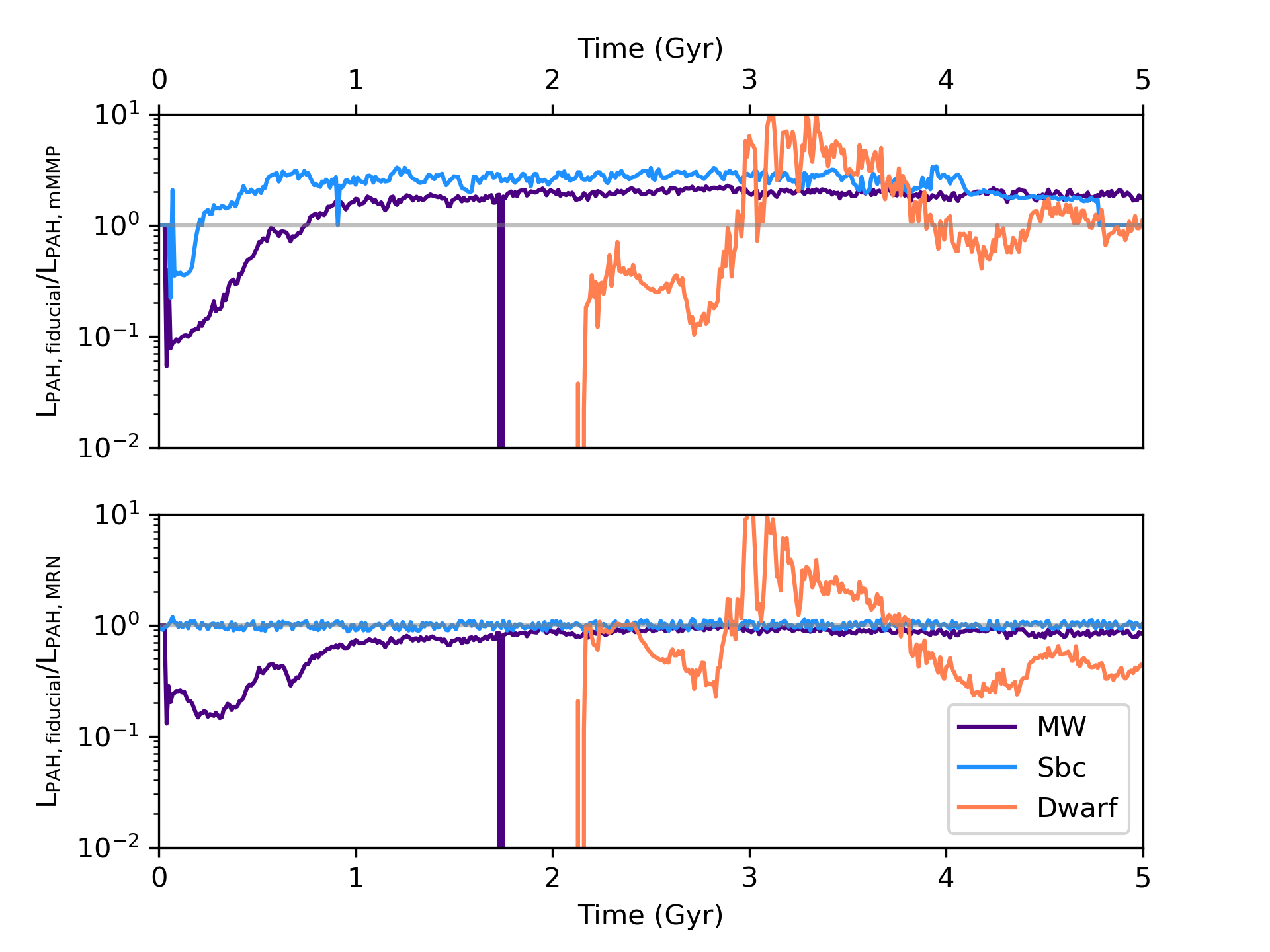}
      \caption{{\bf Numerical experiments investigating the role of
          the interstellar radiation field and dust grain size
          distribution in setting the PAH luminosity in our model
          galaxies.}  {\it Top:} Ratio of \lpah \ evolution from our
        fiducial model to one in which we assume a constant
        \citet{mathis83a} Solar neighborhood-like radiation field
        spectral shape.  {\it Bottom:} Ratio of \lpah \ evolution from
        our fiducial model to one in which we assume a constant
        \citet{mathis77a} MRN-like dust size distribution everywhere
        in each galaxy.  In both plots the solid gray line shows a
        ratio of unity.  The model MW and Sbc vary the most strongly
        from the fiducial run when the radiation field is held fixed,
        as opposed to when the grain size distribution is held fixed.
        This can be interpreted as the hardness of the interstellar
        radiation field impacting the PAH luminosities in these models
        more than the dust grain size distribution.  For the model
        Dwarf galaxy, both the size distribution and ISRF shape are of
        comparable importance.  Future cosmological models will be
        able to address the origin of \lpah \ in galaxies in a
        robust manner.\label{figure:lpah_constrad_constmrn}}
    \end{figure*}

  We now turn to the luminosities of PAHs in our simulated
  galaxies. In Figure~\ref{figure:lpah}, we present the time evolution
  of the bolometric ($\lambda = [3.3,20] \mu$m) PAH luminosities for
  each of our model galaxies.  We show these both normalized by the
  total dust mass (top of Figure~\ref{figure:lpah}), as well as by the
  total far infrared (FIR) luminosity in order to compare to
  observations (bottom of Figure~\ref{figure:lpah}).

Fundamentally, the PAH spectrum from a galaxy depends on: (i) the
total number of PAH dust grains, (ii) the size distribution of those
PAH grains, (iii) the hardness of the radiation field, and (iv) the
intensity of the radiation field.  Variations in $L_{\rm PAH}/M_{\rm
  PAH}$ between models in Figure~\ref{figure:lpah} are therefore ascribed to
either the grain size distribution or properties of the ISRF.  In order to disentangle these effects, we run a
series of controlled numerical experiments in which we compute the PAH
luminosity with {\sc powderday} by (a) assuming a constant
\citet[][mMMP]{mathis83a} Solar neighborhood-like interstellar radiation field shape with the same total bolometric luminosity 
\citep[as modified and distributed by][]{draine21a} though allowing
the grain sizes to vary spatially as computed in the hydrodynamic
galaxy evolution simulation, and by (b) assuming a constant
\citet[][MRN]{mathis77a} Milky Way-like grain size distribution everywhere,
but allowing the ISRF to vary as computed by the stellar population
synthesis and dust radiative transfer in {\sc powderday}.  We present
the results from these experiments in Figure~\ref{figure:lpah_constrad_constmrn}. 

 When comparing the PAH
luminosities from our fiducial model to one in which we assume a
constant mMMP solar-neighborhood like spectral shape
(Figure~\ref{figure:lpah_constrad_constmrn}), we see that the PAH
luminosities are larger by factors $2$--$3$ in the steady state in our
fiducial model for all models.  At the same time, when fixing the
grain size distribution the models MW and Sbc luminosities are relatively unchanged from the
fiducial runs, while the model Dwarf continues to vary dramatically.
This may be interpreted, therefore, as the PAH luminosities of the
Sbc and MW models as being driven primarily by their radiation fields, while
the Dwarf model ascribes comparable importance to the grain size
distribution and number of UV photons.  In detail, this owes to a
harder radiation field in models MW and Sbc than the fiducial mMMP
radiation field.  

Finally, it is worth highlighting the correspondence between the
\lpah \ / $L_{\rm FIR}$ ratios with the observed range in the local
Universe \citep{smith07a}, as demonstrated in the
bottom panel of Figure~\ref{figure:lpah}.  The fraction of total
infrared luminosity that emerges in the PAH bands is a function of
both the fraction of total dust mass that is in the form of PAHs
(\S~\ref{section:qpah}), as well as the total amount of stellar light
that is reprocessed by dust into the infrared.  The strong
correspondence between our models and observations therefore is a
result both of modeled \qpah \ fractions comparable to those observed
in the local Universe, as well as reasonable star/dust geometries in
our modeled disk galaxies.

\subsection{Variations in Individual Feature Strengths}
The mid-IR PAH spectrum is composed of a series of individual
features between $\sim 3.3$--$17$ \micron.  The strengths of these
features vary with interstellar radiation field shape, grain size
distribution, and ionization state \citep{draine21a}.  
While fitting individual
feature strengths is outside the scope of this work,
we briefly present an empirical demonstration of the variation of
individual features in our $3$ model galaxies.

\begin{figure*}
\includegraphics[]{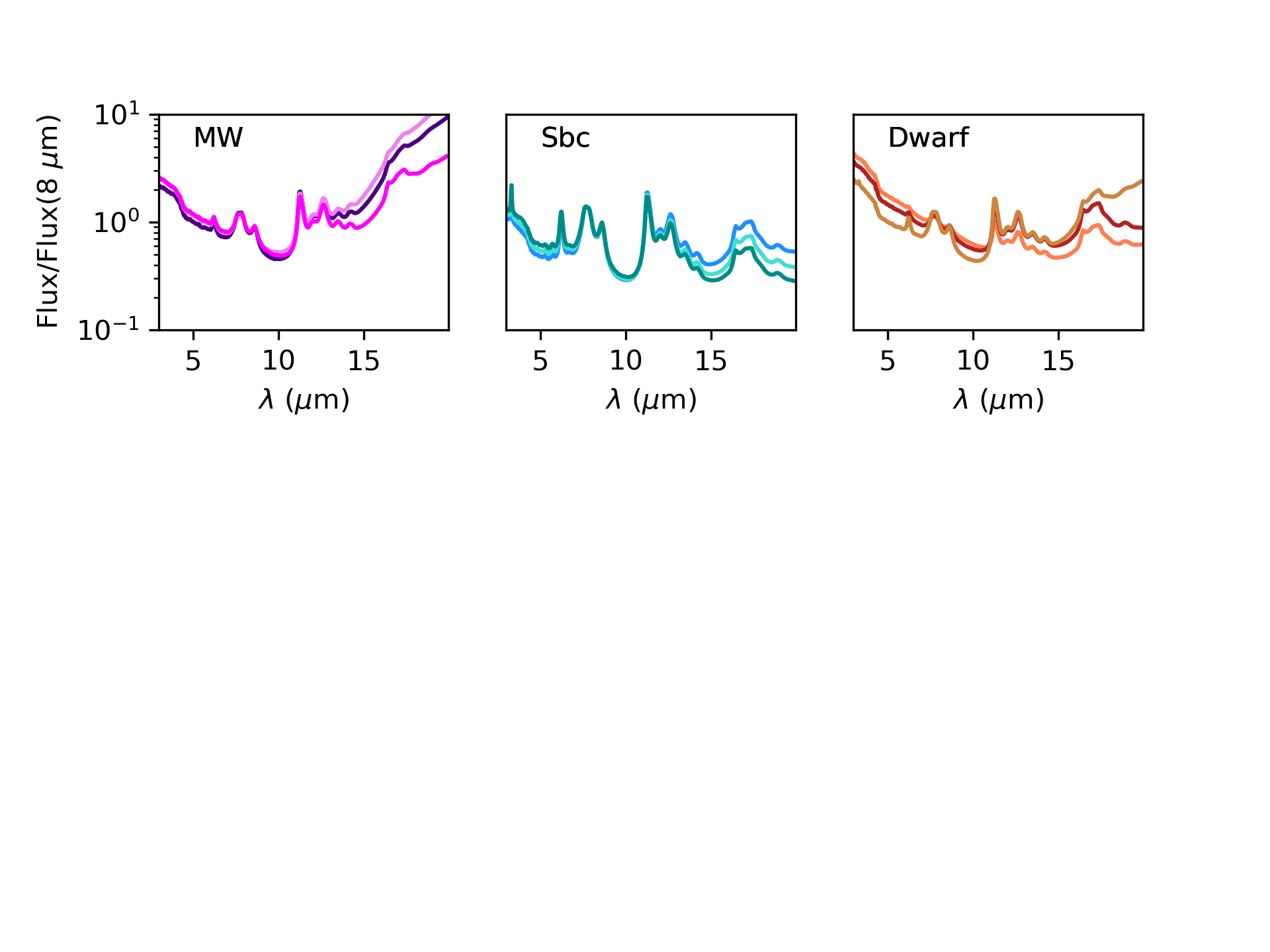}
\vspace{-7cm}
\caption{{\bf Feature strength variations at $3$ arbitrary times
    spanning $2$ Gyr for each of our model galaxies.}  The individual
  colors in each panel denote different evolutionary times, and the
  spectra are normalized at $8 \micron$.  Feature strength variations are common and expected as galaxies
  evolve. \label{figure:feature_strengths}}
\end{figure*}

In Figure~\ref{figure:feature_strengths}, we show the mid-IR SED for
each model galaxy (normalized at $8$ \micron) for each of our galaxies at
$3$ individual time stamps.  In general, we see significant relative
variations in individual features in our model Sbc and Dwarf galaxies,
while the model MW tends to maintain a more constant mid-IR SED during
its evolution.  Referring to Figure~\ref{figure:physical_properties},
this can be understood from the evolution of the physical properties
of the galaxies themselves.  Model MW maintains a fairly constant star
formation history, and consequently a fairly constant grain size distribution due to the
self-regulation of star formation and ISM properties as driven by the
{\sc smuggle} feedback model \citep{marinacci19a}.  At the same time,
the feedback strength (normalized by the galaxy mass) for the models
Sbc and Dwarf are significantly stronger, and drive large
variations in the star formation history (and hence ISRF), as well as grain size distribution\footnote{As a
  reminder, in our model the ionization fraction is tied to the dust
  grain size distribution, and hence the two couple together to drive
  variations in feature strengths}.  The main takeaway is that the
evolution of feature strengths (and their relative ratios) clearly
depends on the evolution of the ISM physical properties within
galaxies, and is unlikely to be captured (in, e.g., SED fitting codes)
from a set of fixed PAH spectral templates. 

\section{Discussion}
\label{section:discussion}
\subsection{Relationship to Other Models}

In developing a model for PAH emission from hydrodynamic galaxy
evolution simulations, we have combined theoretical calculations of
the emission features of stochastically heated dust grains
\citep{draine21a} with a model for the evolution of dust grains with a
range of sizes in galaxy simulations \citep[][Q. Li et al. in
  prep.]{li21a}.  This is, to our knowledge, the first such model, and
it is instructive to place our model into context by discussing both
the range of types of galaxy models that include dust, as well as PAH
emission models.

The most common types of galaxy simulations include dust as a
single-species, single-size passive scalar that is physically tied to
the gas in galaxies.  This has been performed in hydrodynamic galaxy
evolution simulations
\citep[e.g.,][]{asano13a,mckinnon16a,mckinnon17a,aoyama17a,aoyama18a,dave19a,hou19a,hu19a,li19a,vogelsberger19a,graziani20a,esmerian22a},
as well as semi-analytic models that evolve the physical properties of
galaxies in a cosmological context analytically
\citep{popping17b,vijayan19a,triani20a,triani21a}.  \citet{choban22a}
increased the sophistication of these single-size models by including
species-dependent physical processes.

Modeling a full size spectrum of dust grains in galaxy evolution
simulations adds significant computational cost compared to
single-size models. At the same time, understanding galaxy grain size
distributions, extinction laws, and ultimately PAH abundances and
emission, requires an understanding of grain size distribution
variations with galaxy physical properties.  \citet{hirashita15a}
developed a two-size approximation for a means of modeling grain size
evolution, while keeping computational costs reasonable.  This
two-size model, or extensions that include a full size distribution,
has been implemented by a number of groups either on-the-fly in galaxy
evolution evolution simulations, or applied in post-processing
\citep{aoyama18a,gjergo18a,mckinnon18a,hou19a,aoyama20a,huang21a,li21a,romano22a}.

\citet{seok14a}, \citet{hirashita20a}, \citet{hirashita20b} and \citet{hirashita22a}
expanded on the aforementioned works, and developed a formulation for
modeling silicates, aliphatic, and aromatic carbon species in models
with multiple grain size distributions.  These authors assumed that
the smallest aromatic grains were PAHs, and studied the \qpah
\ -\ $Z_{\rm gas}$ relationship in one-zone galaxy models, as well as
the time evolution of PAH abundances. \citet{rau19a} built these
methods into a post-processed model of an idealized isolated disk
galaxy (akin to those we study here) in order to model the evolution
of PAH abundances.  \citet{rau19a} find that the PAH abundances are
driven primarily by shattering and dust growth in galaxies.  In
contrast, \citet{hirashita22a} utilize one-zone models to hypothesize that PAHs do not undergo
significant interstellar processing (i.e., shattering and coagulation),
and find in this scenario a favorable match between their models and
observed PAH abundances, the dust extinction law, and far infrared
SED.  Note in the \citet{hirashita22a} model, PAHs can be formed via
normal channels (i.e., growth and grain-grain shattering) though once formed, do
not undergo interstellar processing.  Our model finds shattering as
the dominant physical process in driving PAH abundances, with coagulation and growth of the smallest dust grains formed in stellar ejecta as negligible.  In this sense, our interpretation of the origin of PAH abundances is different from that of 
the \citet{hirashita22a} one-zone model, though we are not yet in a
position to compare against a diverse range of observational
constraints such as the observed extinction law or $2175 \angstrom$
bump strength.


\subsection{Open Issues and Caveats} 

Our multiscale model for computing the PAH emission spectrum in
hydrodynamic simulations involves tying together models for dust,
radiative transfer, and galaxy simulations over a large dynamic range
of scales, each of which has their own inherent assumptions.  As a
result, our model has, baked into it, a number of open issues and
caveats that we discuss in turn here.

First, there is a series of open issues related to the interface
between the \citet{draine21a} model, and our galaxy
evolution/radiative transfer simulations.  For example, the
\citet{draine21a} model derives the emergent PAH spectrum for a given
grain size distribution and ionization state in terms of the incident
starlight intensity, described by the dimensionless intensity
parameter in Equation~\ref{equation:intensity_parameter}.  As
demonstrated by \citealt{draine21a}, the emergent PAH spectrum between
$\lambda=[3-20] $\micron \ does not vary strongly so long as $U \lesssim 10^3$.
In our model, we do not find many situations where $U \gg 10^3$.  This
said, this is potentially a resolution-dependent phenomenon: it may be
that by being unable to resolve regions in the immediate vicinity of
massive stars, we are missing regions with very large intensity
parameters.  It is unlikely that galaxy-wide (idealized or
cosmological) simulations will be able to achieve this level of
resolution in the near future, though this is an area where
individual ISM patch simulations may perform well
\citep[e.g.][]{walch15a,kim20a}.

Similarly, the \citet{draine21a}
model assumes a carbon-hydrogen (C:H) ratio that evolves with grain
size, as is assumed in \citet{draine07a}.  Because our galaxy
simulations model the chemical composition of dust grains on the fly,
this represents an inconsistency between the underlying
\citet{draine21a} dust model and our galaxy simulations.

Second, we turn our attention to the galaxy evolution simulations
themselves (independent of the \citealt{draine21a} model).  Here, there
are four major open issues related to the interface between dust
grains and the radiation field. (i) Our computation of the far
ultraviolet interstellar radiation field strength when determining
aromitization rates is performed on the fly during the galaxy
evolution simulations by assuming a \citet{mathis83a} radiation field
spectral shape, and a \citet{weingartner01a} dust extinction law
between nearby stars and the dust particle of interest.  Neither of
these are consistent with the true spectral shape\footnote{We remind
the reader, however, that the spectral shape for the UV heating of PAH
grains is calculated explicitly via $3$D dust radiative transfer, in
post-processing.  This radiative transfer takes the spatially resolved
extinction law as determined by the local grain properties into
account. }, or dust extinction law.  (ii) We do not include the
radiative destruction of dust grains.  (iii) The dust grains do not
(currently) impact the models for radiative feedback in the {\sc
  smuggle} simulations \citep{marinacci19a}.  (iv) Our computation of the ionization state of PAHs is parameterized by a relatively simple equation tying the ionization fraction to the grain size distribution (Equation~\ref{equation:ionization}).    This latter issue can significantly impact our modeled feature strength ratios as for a given grain size distribution and ISRF, ionized and neutral PAHs have different mid-IR spectra (e.g. Figure~\ref{figure:sed_decomposed}).
  
  All of these issues are
solvable via the same technique: by connecting our model with a radiation hydrodynamics
solver, we can explicitly compute the impact of the radiation field on dust properties 
\citep[e.g.][]{mckinnon21a}.  
While this is outside the scope of the current modeling effort, future models will merge our dust model with the radiation hydrodynamics branches of the {\sc smuggle} galaxy formation model \citep[e.g.][]{kannan20a}. 
We note, however, that the
computation of the PAH heating rate will still require post-processing
as simulating a sufficiently high spectral resolution in radiation
hydrodynamics is currently computationally intractable.

Finally, a major uncertainty in our model lies in the conversion of aromatic carbonaceous grains to aliphatic ones (and vice versa).   Ultimately, it is unclear if this is even a dominant part of the PAH lifecycle, and whether carbonaceous grains can convert back and forth. The aliphatization rates that we employ are a subresolution model that derive from calculations by \citet{hirashita20a}.  In developing this, \citet{hirashita20a} assume that aliphatization only occurs in dense gas (defined as $n>300$ cm$^{-3}$), which we adopt without any tuning here.  This said, this choice impacts our results: decreasing, or removing this threshold will increase the aliphatization rates and decrease the mass fraction of dust in the form of PAHs ($q_{\rm PAH}$).  Similarly, increasing this threshold will have the opposite effect.  We have not tuned our model based on this aliphatization rate, though alternative implementations of aromitzation and aliphatization would almost certainly impact our methodology and modeled results.   Beyond this, our model assumes that dehydrogenization dominates the aromatization process: the lack of other avenues for aromatization in our model constitutes a major uncertainty.

\section{Summary and Outlook}
\label{section:conclusions}
 We have developed a new framework for
modeling the mid-infrared emission from PAHs in hydrodynamic galaxy
evolution simulations.  We have done this by combining theoretical
single-photon heating models of ultrasmall dust grains with galaxy
evolution simulations that simulate the evolution of a size
distribution of dust grains on-the-fly, as well as the local radiation
field and heating rate.  These new simulations account for the
variation of PAH feature strengths due to grain size distributions,
starlight intensity, and dust composition, and allow us to connect the
evolution of galaxy physical properties to the emergent and varying PAH spectrum.
We describe this new methodology, as well as the relevant equations in
\S~\ref{section:methods}.

We have implemented these methods within the {\sc smuggle} galaxy
formation physics framework, and simulated $3$ idealized disk galaxies
(a Milky Way-like galaxy, a Dwarf, and a Sbc type starburst disk) with
the {\sc arepo} hydrodynamics code in order to investigate the buildup
of PAH masses and luminosities in galaxies.  We describe the {\sc
  smuggle} physics in our galaxy simulations in \S~\ref{section:smuggle}, as well as the
particulars of these idealized galaxies that we employ for numerical
experiments in \S~\ref{section:galaxy_evolution}.  The evolution of
the relevant physical properties for these galaxies is presented in
Figure~\ref{figure:physical_properties}.  We demonstrate an example
model PAH spectrum and image in Figure~\ref{figure:sed_decomposed}.\\

Our main results follow:

\begin{enumerate}

  \item In our model, we allow PAHs to form from both growth of the smallest dust grains, as well as shattering (i.e., interstellar processing) of larger grains. In our model the latter dominates. The key physical processes in driving the formation of
    ultrasmall aromitized carbonaceous dust grains (i.e., PAHs) are
    large velocity dispersions in the ISM (in order to drive
    grain-grain shattering, which pushes the power in the grain size
    distribution toward small grain sizes), and large radiation fields
    (in order to convert aliphatic grains into aromatic
    ones).  This is demonstrated in Figures ~\ref{figure:qpah} and ~\ref{figure:qpah_tuner}.

    \item Increased shattering rates (driven by large ISM velocity
      dispersions) in galaxies are associated with high specific
      star formation rate, which translates to increased feedback
      energy per unit ISM mass.  Aromatization driven by UV radiation
      is accomplished via large global star formation rates.

    \item The fraction of total dust mass that is in the form of PAHs
      ($q_{\rm PAH}$) can be understood as a consequence of these
      processes.  We demonstrate the impact of the shattering
      timescales and aromatic fractions on the modeled \qpah
      \ from our idealized galaxies in Figure~\ref{figure:qpah}.  The
      model starburst has the largest fraction of its dust mass in the
      form of PAHs ($q_{\rm PAH} \approx 0.03-0.2$), the Dwarf the least
      ($q_{\rm PAH} \sim 0.01$), and the model MW in the middle ($q_{\rm PAH} \approx 0.03-0.1$). 

    \item We find that the dominant driver in PAH luminosities in our
      models is the hardness of the interstellar radiation field
      (which translates to the heating rate of PAH dust grains).  This
      is demonstrated in Figures~\ref{figure:lpah} and\ref{figure:lpah_constrad_constmrn}.  That said, this is likely
      a result that is specific to these idealized galaxy models, and
      full cosmological simulations will be necessary to make a
      general statement of the importance of radiation field hardness
      over other contributing factors to the PAH luminosity.  
      
\end{enumerate}

In the era of JWST/MIRI, models such as the ones presented here will complement
mid-infrared observations of PAHs in galaxies near and far.  While
this paper has emphasized the development of a new methodology, we caution that many of the subresolution modeling elements are uncertain.  Our overarching goal has been to establish a framework for modeling PAH emission in galaxy simulations, with a keen eye toward forthcoming JWST observations as a means to constrain and refine the input physics into this model.  As our understanding of PAH physics evolves, individual aspects of this model can be updated.


\section*{Acknowledgements}
 D.N. and JDS express gratitude toward the Aspen Center for Physics
 which is supported by National Science Foundation grant PHY-1607611,
 as well as the organizers of the February 2020 meeting ``Quenching and
 Transformation throughout Cosmic Time'', where the idea for this
 project was borne out of a discussion on a chair lift during a
 wonderful day on Ajax mountain.  D.N. thanks Aaron Evans, Adam
 Ginsburg, Christopher Lovell, Sidney Lower,
 Prerak Garg, Casey Papovich, George Privon, Jia Qi, Julia Roman-Duval, Heath Shipley, and Tom
 Robitaille for helpful conversations.  D.N. additionally thanks Bruce
 Draine for providing early access to the \citet{draine21a}
 models. The authors thank Hiroyuki Hirashita both for helpful conversations during the development of this study, as well as for commenting on an advance draft of this paper.
 JDS gratefully acknowledges support for this project from the Research Corporation for Science Advancement through Cottrell SEED Award No. 27852.  It is a rare agency for which ``Risky, interdisciplinary, and exploratory projects are strongly encouraged.''
 The authors thank T.J. Cox, Phil Hopkins, Brant Robertson and
 Volker Springel for their early work on the initial conditions
 generator for idealized simulations, employed in this study.  D.N.,
 \& P.T. acknowledge support from NASA ATP grant 80NSSC22K0716 and HST-AR-16145.001
 from the Space Telescope Science Institution for funding this work. 
P.T. acknowledges support from NSF grant AST-2008490. The Cosmic Dawn Center is funded by the Danish National Research Foundation under grant No. 140. K.S. acknowledges support from NASA ADAP grant 80NSSC21K0851.



\bibliographystyle{mnras}
\bibliography{./full_refs_pahs}

\end{document}